\documentclass[12pt]{article}
\usepackage{epsfig, epsf, graphicx}
\usepackage{pstricks, pst-node, psfrag}
\usepackage{amssymb,amsmath,bm}
\usepackage{verbatim,enumerate}
\usepackage{diagbox}
\usepackage{rotating, lscape}
\usepackage{setspace}
\usepackage{paralist}
\usepackage{apacite}
\usepackage{natbib}
\usepackage{physics}
\usepackage{color, colortbl}
\usepackage{tabu}
\usepackage{subcaption}
\usepackage{xcolor}
\usepackage{contour}

\definecolor{seablue}{RGB}{0, 103, 225}

\usepackage[framemethod=default]{mdframed}
\usepackage{showexpl}

\mdfdefinestyle{exampledefault}{%
linecolor=black,linewidth=1.5pt,
middlelinewidth=3pt,rightline=true,innerleftmargin=10,innerrightmargin=10}
\usepackage{tikz}
\usepackage{multirow}
\usepackage{bbm}
\usetikzlibrary{arrows,calc,tikzmark,shapes,fit,positioning}
\tikzset{box/.style={draw, rectangle, thick, text centered, minimum height=3em}}
  \tikzset{line/.style={draw, thick, -latex'}}
\usepackage{hyperref}
\usepackage{booktabs}
\usepackage{float}
\usepackage{multicol,listings,siunitx}
\usepackage{caption}
\usepackage{gensymb}
\def\boxit#1{\vbox{\hrule\hbox{\vrule\kern6pt
          \vbox{\kern6pt#1\kern6pt}\kern6pt\vrule}\hrule}}

\def\bse{\begin{eqnarray*}}
\def\ese{\end{eqnarray*}}
\def\be{\begin{eqnarray}}
\def\ee{\end{eqnarray}}
\def\bq{\begin{equation}}
\def\eq{\end{equation}}
\def\bse{\begin{eqnarray*}}
\def\ese{\end{eqnarray*}}

\definecolor{seagreen}{rgb}{0.18,0.55,0.34}
\definecolor{lawngreen}{rgb}{0.49,0.99,0}
\definecolor{lightsalmon}{rgb}{1,0.63,0.48}
\definecolor{lightyellow}{rgb}{0.99,0.906,0.429}

\newtheorem{theo}{Theorem}

\DeclareFontFamily{OT1}{pzc}{}
\DeclareFontShape{OT1}{pzc}{m}{it}{<-> s * [1.10] pzcmi7t}{}
\DeclareMathAlphabet{\mathpzc}{OT1}{pzc}{m}{it}

\begin{document}

\thispagestyle{empty} \baselineskip=28pt \vskip 5mm
\begin{center} {\LARGE{\bf Spatio-Temporal Cross-Covariance Functions under the Lagrangian Framework with Multiple Advections}}
	
\end{center}

\baselineskip=12pt \vskip 10mm

\begin{center}\large
Mary Lai O. Salva\~{n}a, Amanda Lenzi, and Marc G.~Genton\footnote[1]{\baselineskip=10pt Statistics Program, King Abdullah University of Science and Technology (KAUST), Thuwal 23955-6900, Saudi Arabia\\
E-mail: marylai.salvana@kaust.edu.sa amanda.lenzi@kaust.edu.sa marc.genton@kaust.edu.sa\\
This research was supported by the King Abdullah University of Science and Technology (KAUST).}
\end{center}

\baselineskip=17pt \vskip 10mm \centerline{\today} \vskip 15mm

%%%%%%%%%%%%%%%%%%%%%%%%%%%%%%%%%%%%%%%%%%%%%%%%%%%%%%%%%%%%%%%%%%%%%%%%
\begin{center}
{\large{\bf Abstract}}
\end{center} 

When analyzing the spatio-temporal dependence in most environmental and earth sciences variables such as pollutant concentrations at different levels of the atmosphere, a special property is observed: the covariances and cross-covariances are stronger in certain directions. This property is attributed to the presence of natural forces, such as wind, which cause the transport and dispersion of these variables. This spatio-temporal dynamics prompted the use of the Lagrangian reference frame alongside any Gaussian spatio-temporal geostatistical model. Under this modeling framework, a whole new class was birthed and was known as the class of spatio-temporal covariance functions under the Lagrangian framework, with several developments already established in the univariate setting, in both stationary and nonstationary formulations, but less so in the multivariate case. Despite the many advances in this modeling approach, efforts have yet to be directed to probing the case for the use of multiple advections, especially when several variables are involved. Accounting for multiple advections would make the Lagrangian framework a more viable approach in modeling realistic multivariate transport scenarios. In this work, we establish a class of Lagrangian spatio-temporal cross-covariance functions with multiple advections, study its properties, and demonstrate its use on a bivariate pollutant dataset of particulate matter in Saudi Arabia. 

\par\vfill\noindent
{\bf Some key words:} cross-covariance function; Lagrangian framework; multiple advections; multivariate random field; spatio-temporal; transport effect.

\clearpage\pagebreak\newpage \pagenumbering{arabic}
\baselineskip=26pt

%%%%%%%%%%%%%%%%%%%%%%%%%%%%%%%%%%%%%%%%%%%%%%%%%%%%%%%%%%%%%%%%%%%%%%%%

\section{Introduction}\label{sec:intro}

Many environmental and earth sciences datasets record several variables at certain locations over certain periods of time. These datasets contain rich spatio-temporal information which can be used to enhance predictions. In geostatistics, spatio-temporal data are used to calibrate the parameters of spatio-temporal cross-covariance functions which are valid functions that describe the spatio-temporal relationships within each variable and between any two variables; see \cite{genton2015cross}, \cite{alegria2019covariance}, and \cite{salvana2020nonstationary} for recent reviews of the available models in the literature. In multivariate Gaussian spatio-temporal geostatistical modeling, we work with a spatio-temporal process 
$\mathbf{Y}(\mathbf{s},t)=\left\{Y_1(\mathbf{s},t),\ldots,Y_p(\mathbf{s},t)\right\}^{\top}$, $(\mathbf{s},t)\in  \mathbb{R}^d \times \mathbb{R}$,
such that at each spatial location $\mathbf{s} \in \mathbb{R}^d, d\geq 1$, and at each time $t\in\mathbb{R}$, there are $p$ variables. A common assumption on $\mathbf{Y}(\mathbf{s},t)$ is that it can be decomposed into a sum of a deterministic and a random component, i.e.,
 \begin{equation} \label{eqn:linear_model}
\mathbf{Y}(\mathbf{s},t) =\boldsymbol{\mu}(\mathbf{s},t) + \mathbf{Z}(\mathbf{s},t),
\end{equation}
where $\boldsymbol{\mu}(\mathbf{s}, t)$ is a mean function and $\mathbf{Z}(\mathbf{s}, t)$ is a zero-mean multivariate Gaussian spatio-temporal process. When $\mathbf{Z}(\mathbf{s}, t)$ is second-order stationary, it is completely characterized by its spatio-temporal matrix-valued stationary cross-covariance function $\mathbf{C}(\mathbf{h},u) = \left\{C_{ij}(\mathbf{h},u)\right\}_{i,j=1}^{p}$ on $\mathbb{R}^d \times \mathbb{R}$, with entries defined as follows:
\begin{equation} \label{eqn:stationarity}
C_{ij}(\mathbf{h},u)=\text{cov}\left\{Z_i(\mathbf{s},t),Z_j(\mathbf{s}+\mathbf{h},t+u)\right\}, \quad (\mathbf{h},u) \in \mathbb{R}^{d}\times \mathbb{R},
\end{equation}
for $i,j=1,\ldots,p$. Often, $C_{ij}$ is termed the marginal covariance function when $i = j$ and is called the cross-covariance function whenever $i \neq j$. From (\ref{eqn:stationarity}), it is easy to check that $C_{ij}(\mathbf{h},u) = C_{ji}(-\mathbf{h},-u)$, for any $(\mathbf{h},u) \in \mathbb{R}^{d}\times \mathbb{R}$ and $i, j = 1, \ldots, p$. However, it is not always the case that $C_{ij}(\mathbf{h},u) = C_{ij}(\mathbf{h}, -u) = C_{ij}(-\mathbf{h}, u) = C_{ij}(-\mathbf{h},-u)$ for $i, j = 1, \ldots, p$. This equality involving different combinations of the signs of the spatial lag, $\mathbf{h}$, and temporal lag, $u$, is referred to as the full spatio-temporal symmetry property. A consequence of the full spatio-temporal symmetry property is that the marginal and cross-covariances between observations at $(\mathbf{s}, t)$ and $(\mathbf{s} + \mathbf{h}, t + u)$ are equal to the marginal and cross-covariances between observations at $(\mathbf{s}, t + u)$ and $(\mathbf{s} + \mathbf{h}, t)$. This property is highly restrictive and most often not exhibited by environmental and earth sciences data. For instance, when wind blows from West to East, an airborne substance can be transported such that its concentration at a certain site will be highly correlated to the concentration, measured after some time, at a site to its East. Because of the specific direction of transport, the same degree of correlation may not be expressed between the concentration at the aforementioned site and the concentration, measured after some time, at a site to its West. This behavior is recognized as spatio-temporal asymmetry in the marginal and cross-covariance structures \citep{gneiting2006geostatistical, huang2020visualization, chen2021space}. Whenever such asymmetry is detected, more appropriate spatio-temporal asymmetric cross-covariance functions ought to be used such as the class of spatio-temporal asymmetric models proposed in \cite{stein2005space}, suitable for $p = 1$, and the latent dimensions model in \citet{apanasovich2010cross}, catering to $p > 1$.

%Despite the ubiquity of environmental phenomena with spatio-temporal asymmetric dependencies, models that accommodate asymmetry in the spatio-temporal dependence structure within the same variable and between any two variables are scarce. 

Within the category of models that captures spatio-temporal asymmetric dependence, a rich subclass was established and is dedicated to transported purely spatial random fields. The models under this subclass are termed spatio-temporal cross-covariance functions under the Lagrangian framework and the pioneering work in the univariate setting is attributed to \citet{cox1988}. Under this modeling framework, consider a purely spatial random field with a stationary covariance function $C^S(\mathbf{h})$ and suppose that this entire random field moves forward in time, with a random advection velocity $\mathbf{V} \in \mathbb{R}^d$. This means that the observations from all spatial locations are being advected or transported by one and the same random advection velocity at each time. The resulting Lagrangian spatio-temporal covariance function has the form:
\begin{equation}\label{eqn:coxishameq}
C(\mathbf{h},u) = \text{E}_{\mathbf{V}}\left\{C^S(\mathbf{h} - \mathbf{V} u )\right\},
\end{equation}
where the expectation is taken with respect to $\mathbf{V}$. Through $\mathbf{V}$, one can transform a purely spatial random field into a spatio-temporal random field. Moreover, by such a transformation to the spatial coordinates that incorporates information regarding a transport phenomenon, a spatio-temporal covariance function can be derived from a purely spatial covariance function. Depending on the distribution assumed by $\mathbf{V}$, the model in (\ref{eqn:coxishameq}) may not have an explicit form, however, numerical solutions can be easily obtained. The simplest form of (\ref{eqn:coxishameq}) may be derived when $\mathbf{V}$ is chosen to be constant, i.e., $\mathbf{V}=\mathbf{v}$, and the model is termed the frozen field model. Several authors have used the frozen field model to analyze wind speeds \citep{gneiting2006geostatistical, ezzat2018spatio} and solar irradiance \citep{lonij2013intra, inoue2012spatio, shinozaki2016areal}. Intuitively, the frozen field model is an unrealistic assumption and a highly idealized model of the transport phenomenon. Wind, for example, may or may not blow at any time of the day, and when it blows, the wind speeds and directions are rarely identical. Therefore, models that allow for variability of the transport, i.e., the non-frozen field models, are preferred. An explicit form of (\ref{eqn:coxishameq}) exists when $\mathbf{V} \sim \mathcal{N}_d (\boldsymbol{\mu}_{\mathbf{V}}, \boldsymbol{\Sigma}_{\mathbf{V}})$ and $C^S(\| \mathbf{h} \|)$ is a normal scale-mixture model \citep{schlather2010some}. The derived non-frozen field model has the form:
\begin{equation} \label{eqn:schlather_nonfrozen}
C(\mathbf{h}, u) = \frac{1}{\sqrt{|\mathbf{I}_d + \boldsymbol{\Sigma}_{\mathbf{V}}u^2|}} C^S \{ \left(\mathbf{h} - \boldsymbol{\mu}_{\mathbf{V}}u\right)^{\top} \left(\mathbf{I}_d+\boldsymbol{\Sigma}_{\mathbf{V}}u^2\right)^{-1} \left(\mathbf{h}- \boldsymbol{\mu}_{\mathbf{V}} u\right) \},
\end{equation}
where $\mathbf{I}_d$ is the $d \times d$ dimensional identity matrix.

Several versions of the Lagrangian spatio-temporal covariance functions have appeared since the seminal work of \citet{cox1988}. \cite{park2006new} adapted the modeling framework for axial symmetry in time, axial symmetry in space, and diagonal symmetry in space. \cite{porcu2006nonseparable} explored some anisotropic extensions and \cite{christakos2017spatiotemporal} introduced an acceleration component. \cite{salvana2020nonstationary} proposed the multivariate extension such that the model in (\ref{eqn:coxishameq}) remains valid when the underlying purely spatial covariance function is a matrix-valued nonstationary cross-covariance function $\mathbf{C}^{S}(\mathbf{s}_1, \mathbf{s}_2)$ on $\mathbb{R}^d$. Their spatio-temporal matrix-valued nonstationary extension has the form:
\begin{equation}  \label{eqn:spatial_stat_proposed}
\mathbf{C}(\mathbf{s}_1, \mathbf{s}_2; t_1, t_2) = \text{E}_{\mathbf{V}} \{ \mathbf{C}^S(\mathbf{s}_1 - \mathbf{V} t_1,\mathbf{s}_2 - \mathbf{V} t_2 ) \}. 
\end{equation} 
The above model includes the purely spatial matrix-valued stationary cross-covariance functions, i.e., $\mathbf{C}^S(\mathbf{h})$, as the underlying purely spatial cross-covariance functions. Their umbrella theorem relies on a single $\mathbf{V}$ which implies that every component of $\mathbf{Z}(\mathbf{s}, t) \in \mathbb{R}^p$ is transported by the same advection velocity. However, different variables may experience different transport patterns which render the model in (\ref{eqn:spatial_stat_proposed}) inadequate. To deal with this issue, they proposed a Lagrangian spatio-temporal cross-covariance function that is a linear combination of uncorrelated univariate Lagrangian spatio-temporal covariance functions, each depending on different advections. Their proposal is a good first step to addressing this multiple advections problem. When the marginal and cross-advections are introduced, i.e., $\mathbf{V}_{ij} \in \mathbb{R}^d, i, j = 1, \ldots, p$, several questions arise regarding the validity of the extended model, including which values of $\mathbf{V}_{ij}$, $i \neq j$, will preserve the positive definiteness of the cross-covariance matrix resulting from (\ref{eqn:spatial_stat_proposed}). In this work, we aim at answering such a fundamental question and providing a comprehensive treatment to the Lagrangian spatio-temporal cross-covariance functions with multiple advections, with a main focus on underlying purely spatial cross-covariance functions that are stationary.

The remainder of the paper is organized as follows. Section \ref{sec:proposed} presents the proposed extension of (\ref{eqn:spatial_stat_proposed}) with multiple advections and introduces some examples. Section \ref{sec:estimation} details the estimation procedure. Section \ref{sec:simulation} investigates the consequences of neglecting multiple advections in multivariate Lagrangian spatio-temporal modeling. Section \ref{sec:application} compares the performance of the proposed models with other benchmark models in the literature using bivariate pollutant data in Saudi Arabia. The conclusion is presented in Section \ref{sec:discussion} and proofs are collected in the Appendix.

\section{Lagrangian Framework with Multiple Advections}\label{sec:proposed}

The validity of Lagrangian spatio-temporal cross-covariance functions with a different advection for each variable can be established by considering a zero-mean multivariate spatio-temporal random field 
\begin{equation} \label{eqn:process_with_multiple_advec}
\mathbf{Z}(\mathbf{s},t)=\big\{\tilde{Z}_1(\mathbf{s} - \mathbf{V}_{11} t), \ldots, \tilde{Z}_p(\mathbf{s} - \mathbf{V}_{pp}t)\big\}^{\top},
\end{equation} 
such that $\tilde{\mathbf{Z}}(\mathbf{s}) = \big\{\tilde{Z}_1(\mathbf{s}), \ldots, \tilde{Z}_p(\mathbf{s} )\big\}^{\top}$ is a zero-mean multivariate purely spatial random field and every component of $\tilde{\mathbf{Z}}$ is transported by different random advections $\mathbf{V}_{ii} \in \mathbb{R}^d, i = 1, \ldots, p$. The resulting matrix-valued spatio-temporal cross-covariance function of the process in (\ref{eqn:process_with_multiple_advec}) is given in the following theorem.
\begin{theo} \label{theorem1}
Let $\mathbf{V}_{11}, \mathbf{V}_{22}, \ldots, \mathbf{V}_{pp}$ be random vectors on $\mathbb{R}^d$. If $\mathbf{C}^S(\mathbf{h})$ is a valid purely spatial matrix-valued stationary cross-covariance function on $\mathbb{R}^d$, then
\begin{equation} \label{eqn:proposed_multiple_advec}
\mathbf{C}(\mathbf{h}; t_1, t_2) = \normalfont \text{E}_{\boldsymbol{\mathcal{V}}}[ \{C_{ij}^S(\mathbf{h} - \mathbf{V}_{ii} t_1 +  \mathbf{V}_{jj} t_2)\}_{i,j=1}^{p}], 
\end{equation}
where the expectation is taken with respect to the joint distribution of $\boldsymbol{\mathcal{V}} = (\mathbf{V}_{11}^{\top}, \mathbf{V}_{22}^{\top}, \ldots, \mathbf{V}_{pp}^{\top})^{\top}$, is a valid matrix-valued spatio-temporal cross-covariance function on $\mathbb{R}^d \times \mathbb{R}$ provided that the expectation exists. 
\end{theo} 
When $\mathbf{V}_{ii} = \mathbf{V}$, for all $i$, the above model reduces to the single advection case. Moreover, the model in (\ref{eqn:proposed_multiple_advec}) can be rewritten to resemble the form in (\ref{eqn:coxishameq}) such that the temporal lag, $u$, appears: 
\begin{equation} \label{eqn:locally_stationary}
\mathbf{C}(\mathbf{h};t_1,t_2) = \text{E}_{\boldsymbol{\mathcal{V}}} ([C_{ij}^{S} \{\mathbf{h} - \overline{\mathbf{V}}_{ij} u + (\mathbf{V}_{jj}-\mathbf{V}_{ii})m \}_{i,j=1}^{p} ]),
\end{equation}
where $\overline{\mathbf{V}}_{ij}=\frac{\mathbf{V}_{ii}+\mathbf{V}_{jj}}{2}$ and $m=\frac{t_1+t_2}{2}$. When $i = j$, the term that depends on the midpoint between $t_1$ and $t_2$ disappears. Hence, the marginal covariances remain stationary in time. When $i \neq j$, nonstationarity in time is introduced by the Lagrangian shift in the cross-covariances. This reveals an important advantage of using the multiple advections model in (\ref{eqn:locally_stationary}). That is, it permits the colocated correlation between two variables, i.e., $C_{ij} (\mathbf{0}; t, t) / \sqrt{C_{ii}(\mathbf{0}, 0) C_{jj}(\mathbf{0}, 0) }$, to change in time, unlike the prevailing spatio-temporal cross-covariance functions in the literature such as \cite{bourotte2016flexible} and \cite{salvana2020nonstationary}. 

An explicit form of (\ref{eqn:proposed_multiple_advec}) can also be derived similar to (\ref{eqn:schlather_nonfrozen}) and is given in the following theorem.

\begin{theo} \label{theorem_multiple_advec_nonfrozen}
For $p \geq 2$, let $\boldsymbol{\mathcal{V}} = (\mathbf{V}_{11}^{\top}, \mathbf{V}_{22}^{\top}, \ldots, \mathbf{V}_{pp}^{\top})^{\top} \sim \mathcal{N}_{pd} (\boldsymbol{\mu}_{\boldsymbol{\mathcal{V}}}, \boldsymbol{\Sigma}_{\boldsymbol{\mathcal{V}}} )$. If $\mathbf{C}^S(\|\mathbf{h}\|)$ is a matrix-valued normal scale-mixture cross-covariance function, then
\begin{eqnarray} \label{eqn:proposed-nonfrozen-multi-advec-marginal}
C_{ii}(\mathbf{h}, u) = \frac{C_{ii}^S \{ (\mathbf{h} - \mathbf{B}_{(d i - 1):( di )}^{\top} \boldsymbol{\mu}_{\boldsymbol{\mathcal{V}}} u )^{\top} (\mathbf{I}_d + \mathbf{B}_{(d i - 1):( di )}^{\top} \boldsymbol{\Sigma}_{\boldsymbol{\mathcal{V}}} u^2)^{-1} (\mathbf{h} - \mathbf{B}_{( di - 1):( di )}^{\top}\boldsymbol{\mu}_{\boldsymbol{\mathcal{V}}} u ) \}}{|\mathbf{I}_d + \mathbf{B}_{(d i - 1):( di )}^{\top} \boldsymbol{\Sigma}_{\boldsymbol{\mathcal{V}}} u^2|^{1/2}} ,
\end{eqnarray}
where $ \mathbf{B}_{( di - 1):( di )}$ is the sub-matrix of $\mathbf{I}_{pd}$, comprised of its $( di - 1)$-th and $( di )$-th rows, for $i = 1, \ldots, p$, and 
\begin{eqnarray} \label{eqn:proposed-nonfrozen-multi-advec-cross}
C_{ij} (\mathbf{h}; t_1, t_2) = \frac{C_{ij}^S ( (\mathbf{h} - \mathbf{T} \tilde{\mathbf{B}}^{\top} \boldsymbol{\mu}_{\boldsymbol{\mathcal{V}}})^{\top} [ \mathbf{I}_d - \mathbf{T} \{ \mathbf{T}^{\top} \mathbf{T} + ( \tilde{\mathbf{B}}^{\top} \boldsymbol{\Sigma}_{\boldsymbol{\mathcal{V}}})^{-1} \}^{-1} \mathbf{T}^{\top} ] (\mathbf{h} - \mathbf{T} \tilde{\mathbf{B}}^{\top} \boldsymbol{\mu}_{\boldsymbol{\mathcal{V}}})  )  }{{ |  \mathbf{I}_{2 d} +  (\tilde{\mathbf{B}}^{\top} \boldsymbol{\Sigma}_{\boldsymbol{\mathcal{V}}}) \mathbf{T}^{\top} \mathbf{T} |^{1/2}}},
\end{eqnarray}
where $\mathbf{T} = (t_1 \mathbf{I}_d \quad -t_2 \mathbf{I}_d )$, $\tilde{\mathbf{B}} = \mathbf{B}_{\{(d i - 1):( di ), (d j - 1):( dj )\}}$, such that $\mathbf{B}_{\{(d i - 1):( di ), (d j - 1):( dj )\}}$ is the sub-matrix of $\mathbf{I}_{pd}$ comprised of its $( di - 1)$-th, $( di )$-th, $( dj - 1)$-th, and $( dj )$-th rows, for $i, j = 1, \ldots, p$, $i \neq j$.
\end{theo}

Several properties of non-frozen Lagrangian spatio-temporal cross-covariance functions with multiple advections can be identified based on Theorem \ref{theorem_multiple_advec_nonfrozen}. First, the spatial lag at which the maximum value occurs is at $\mathbf{B}_{(d i - 1):( di )}^{\top} \boldsymbol{\mu}_{\boldsymbol{\mathcal{V}}} u $, for the marginals, and at $\mathbf{T} \tilde{\mathbf{B}}^{\top} \boldsymbol{\mu}_{\boldsymbol{\mathcal{V}}}$, for the cross-covariances. This means that for a given temporal lag $u$, variable $i$ measured at location $\mathbf{s}$ is highly dependent with variable $i$ measured at a location $\mathbf{B}_{(d i - 1):( di )}^{\top} \boldsymbol{\mu}_{\boldsymbol{\mathcal{V}}} u $ away from $\mathbf{s}$, $i = 1, \ldots,p$. Moreover, for a given $t_1$ and $t_2$ pair, variable $i$ at $\mathbf{s}$ is highly dependent with variable $j$ found at location $\mathbf{T} \tilde{\mathbf{B}}^{\top} \boldsymbol{\mu}_{\boldsymbol{\mathcal{V}}}$, for $i, j = 1, \ldots, p$ and $i \neq j$. Second, the maximum absolute colocated correlation, $ \max_{\substack{t}} \big| C_{ij} (\mathbf{0}; t, t) / \sqrt{C_{ii}(\mathbf{0}, 0) C_{jj}(\mathbf{0}, 0) } \big|$, occurs at $t = 0$. Third,  the absolute colocated correlation, $\big| C_{ij} (\mathbf{0}; t, t) / \sqrt{C_{ii}(\mathbf{0}, 0) C_{jj}(\mathbf{0}, 0) } \big|$, goes to zero as $|t| \rightarrow \infty$. This means that the two variables at the same location become statistically independent as time moves away from zero.

\begin{figure}[t!]
 \centering
	\includegraphics[scale=0.48]{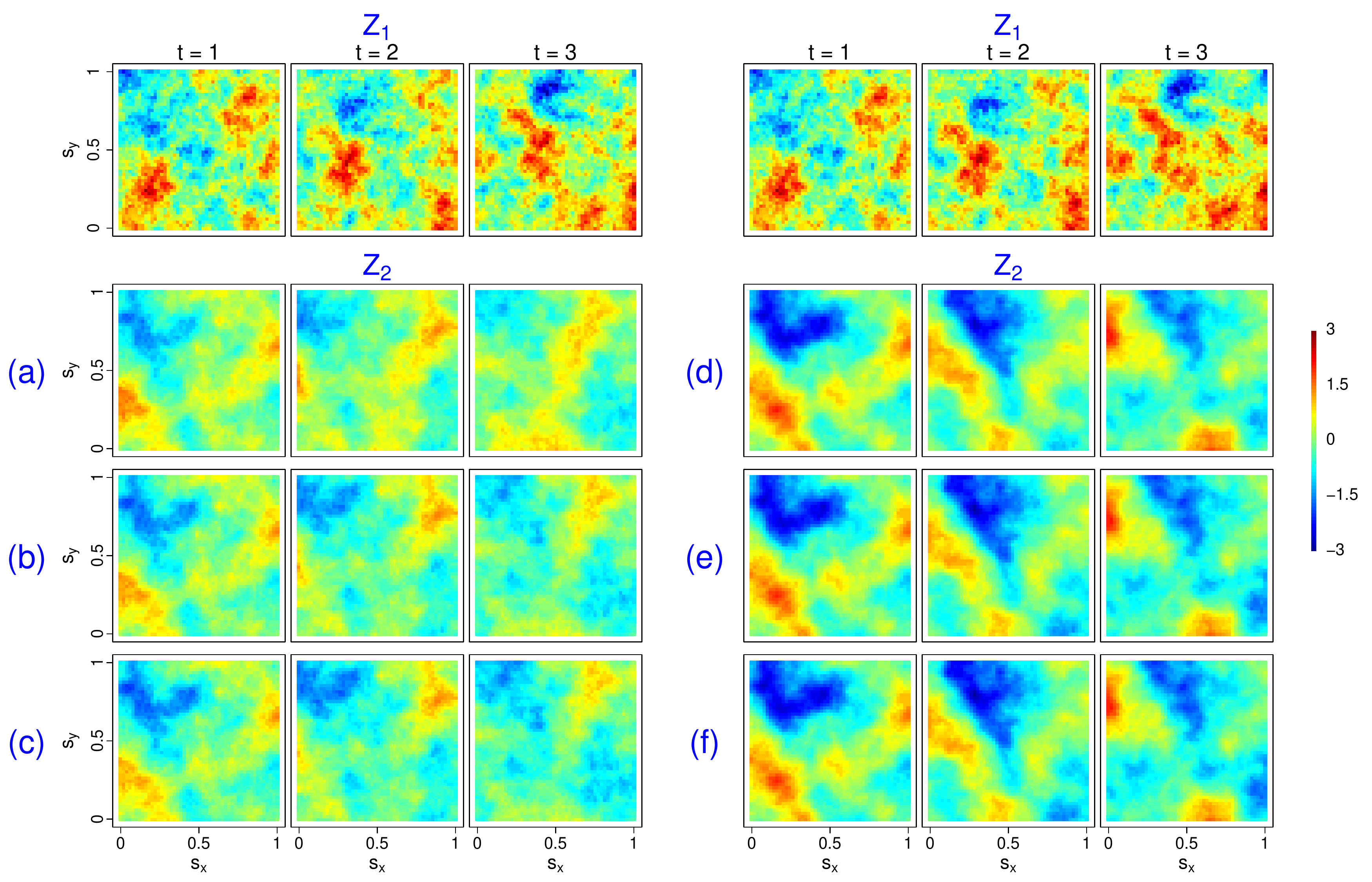}
	\caption{ Simulated realizations of the Lagrangian spatio-temporal parsimonious Mat\'{e}rn cross-covariance function for $p=2$ and $d = 2$, on a $50\times 50$ regular grid in the unit square $[0,1]^2$ with purely spatial parameters, namely, $\nu_{11}=0.5,\;\nu_{22}=1.5,\;a=0.23,\;\rho=0.5,\;\sigma_{11}^2=\sigma_{22}^2=1$. The plots on the left hand side show the bivariate spatio-temporal random fields simulated from (\ref{eqn:proposed-nonfrozen-multi-advec-marginal}) and (\ref{eqn:proposed-nonfrozen-multi-advec-cross}), where a common $Z_1(\mathbf{s}, t)$ is simulated for every configuration shown in Panels (a)-(c). The different realizations of $Z_2(\mathbf{s}, t)$ under varying degrees of dependence between $\mathbf{V}_{11}$ and $\mathbf{V}_{22}$, namely, (a) $\mathbf{V}_{11} = 0.9 \mathbf{V}_{22}$, (b) $\mathbf{V}_{11}$ and $\mathbf{V}_{22}$ are independent, and (c) $\mathbf{V}_{11} = - 0.9 \mathbf{V}_{22}$ are displayed in Panels (a)-(c). The plots on the right hand side show the bivariate spatio-temporal random fields simulated from (\ref{eqn:proposed_variable_specific_advec}) with $d' = 1$, $\mathbf{s}'_{11} = 0.2$, $\mathbf{s}'_{22} = 0$. Similarly, a common $Z_1(\mathbf{s}, t)$ is simulated for different realizations of $Z_2(\mathbf{s}, t)$ in Panels (d)-(f), where (d) $\text{cov}(\text{V}'_{11}, \text{V}'_{22}) = 0.9$, (e) $\text{cov}(\text{V}'_{11}, \text{V}'_{22}) = 0$, and (f) $\text{cov}(\text{V}'_{11}, \text{V}'_{22}) = -0.9$.}
    \label{fig:fig1}
\end{figure}

Figure~\ref{fig:fig1} shows simulated Lagrangian spatio-temporal bivariate random fields from (\ref{eqn:proposed-nonfrozen-multi-advec-marginal}) and (\ref{eqn:proposed-nonfrozen-multi-advec-cross}) on a $50\times 50$ regular grid in the unit square $[0,1]^2$ for $p = 2$ and $d = 2$, with the parsimonious Mat\'{e}rn cross-covariance function as $\mathbf{C}^S(\|\mathbf{h}\|)$ \citep{gneiting2010matern}. Panels (a)-(c) correspond to three different joint distributions of $\mathbf{V}_{11}$ and $\mathbf{V}_{22}$, namely, (a) $\mathbf{V}_{11} = 0.9 \mathbf{V}_{22}$, (b) $\mathbf{V}_{11}$ and $\mathbf{V}_{22}$ are independent, and (c) $\mathbf{V}_{11} = - 0.9 \mathbf{V}_{22}$. The purely spatial parameters were chosen such that the practical spatial range of the variable with a less smooth field is equal to $0.7$, i.e., $C_{11}(\mathbf{h},0)/C_{11}(\mathbf{0},0)\approx 0.05$ when $\|\mathbf{h}\|=0.7$. The marginal mean advection parameters are as follows: $\text{E}(\mathbf{V}_{11}) = (0.1, 0.1)^{\top}$, $\text{E}(\mathbf{V}_{22}) = (-0.1, 0.1)^{\top}$, and $\text{var}(\mathbf{V}_{11}) = \text{var}(\mathbf{V}_{22}) = 0.1 \times \mathbf{I}_2$.

\begin{figure}[t!]
 \centering
	\includegraphics[scale=0.25]{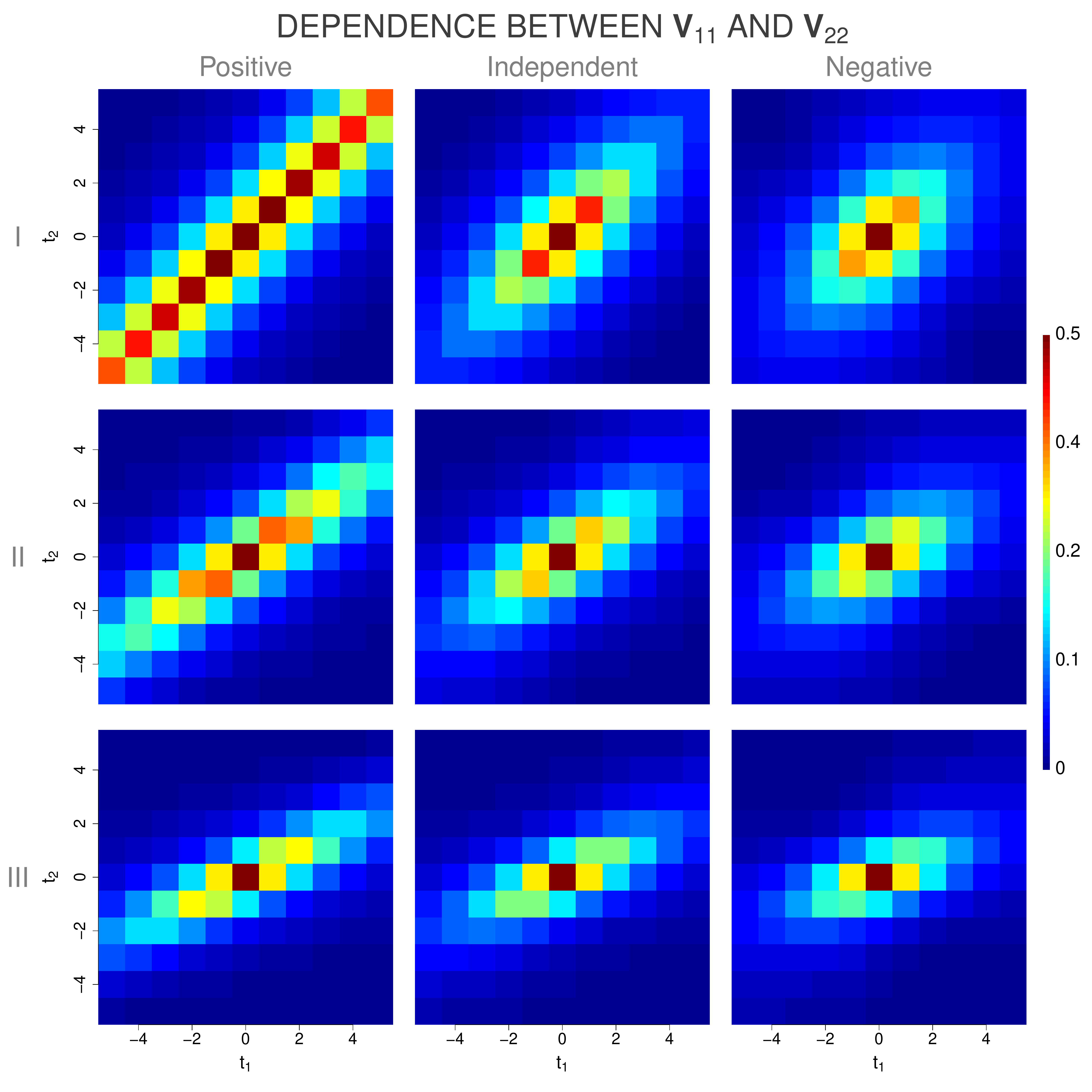}
	\caption{Heatmaps of values of $ C_{12} (\mathbf{0}; t_1, t_2) / \sqrt{C_{11}(\mathbf{0}, 0) C_{22}(\mathbf{0}, 0) } $ for different combinations of temporal locations $t_1 \in \mathbb{Z}$ and $t_2 \in \mathbb{Z}$, $t_1, t_2 \in [-5, 5]$. The purely spatial parameters are set as follows: $\nu_{11}=0.5,\;\nu_{22}=1.5,\;a=0.23,\;\rho=0.5,\;\sigma_{11}^2=\sigma_{22}^2=1$. For all plots, $\text{E}(\mathbf{V}_{11}) = (0.1, 0.1)^{\top}$. Rows I-III correspond to $\text{E}(\mathbf{V}_{22}) = (0.1, 0.1)^{\top}$, $\text{E}(\mathbf{V}_{22}) = (0.2, 0.1)^{\top}$, and $\text{E}(\mathbf{V}_{22}) = (0.3, 0.1)^{\top}$, respectively. The dependence between $\mathbf{V}_{11}$ and $\mathbf{V}_{22}$ in the left, center, and right columns are $\mathbf{V}_{11} = 0.9 \mathbf{V}_{22}$, independent, and $\mathbf{V}_{11} = - 0.9 \mathbf{V}_{22}$, respectively. The evolution of the spatio-temporal dependence between $Z_1$ and $Z_2$ differs depending on the joint distribution of the advection velocities.}
    \label{fig:fig2}
\end{figure}

The three scenarios described above imply different strengths of spatio-temporal dependence. For ease of comparison, we simulate the same $Z_1(\mathbf{s}, t)$ for every configuration and contrast different simulated $Z_2 (\mathbf{s}, t)$. It can be seen in Panels (a)-(c) that while the direction of transport of $Z_2(\mathbf{s}, t)$ is to the North West in all three cases, the spatio-temporal random fields are substantially different, with the different scenarios in the joint distribution of $\mathbf{V}_{11}$ and $\mathbf{V}_{22}$ having visible consequences in the values of $Z_2 (\mathbf{s}, t)$ as time progresses. %It should be noted that scenario (a) is the closest to the single advection case, wherein $\mathbf{V}_{11} = \mathbf{V}_{22}$. 

Figure~\ref{fig:fig2} illustrates how the multiple advections model depart from the single advection model by showing the purely spatial correlation values for different combinations of temporal locations $t_1$ and $t_2$ via heatmaps under nine different advection scenarios. The plots in Figure~\ref{fig:fig2} narrate how strong the dependence is between $Z_1$ and $Z_2$ taken at the same spatial location but at different points in time. That is, $Z_1$, measured at time $t_1$, is either behind, ahead, or at the same time as $Z_2$, measured at time $t_2$. We evaluate the formulas in (\ref{eqn:proposed-nonfrozen-multi-advec-cross}) with different assumptions on $\mathbf{V}_{11}$ and $\mathbf{V}_{22}$. The first row (I) gives the values of the colocated correlation when $\text{E}(\mathbf{V}_{11}) = \text{E}(\mathbf{V}_{22}) = (0.1, 0.1)^{\top}$. The second row (II) shows the values of the colocated correlation when $\text{E}(\mathbf{V}_{11}) = (0.1, 0.1)^{\top}$ and $\text{E}(\mathbf{V}_{22}) = (0.2, 0.1)^{\top}$. The last row (III) presents the values of the colocated correlation when $\text{E}(\mathbf{V}_{11}) = (0.1, 0.1)^{\top}$ and $\text{E}(\mathbf{V}_{22}) = (0.3, 0.1)^{\top}$. Furthermore, the vectors $\mathbf{V}_{11}$ and $\mathbf{V}_{22}$ are assumed to be positively correlated in the first column, i.e., $\mathbf{V}_{11} = 0.9 \mathbf{V}_{22}$, independent in the second column, and negatively correlated in the last column, i.e., $\mathbf{V}_{11} = - 0.9 \mathbf{V}_{22}$. The nonstationarity in time is revealed by the changing correlation values along each diagonal. The single advection model is less flexible as it does not allow for this nonstationarity in the correlation. A stationary in time model will have a constant value along every diagonal. The plot in the $(1,1)$ position in Figure~\ref{fig:fig2} has correlation values that are the closest to the single advection model where the values on the main diagonal should be constant and equal to $\rho$. Even though this particular configuration comes close to the single advection scenario, the nuances appear as $t$, i.e., $t = t_1 = t_2$, moves away from zero. When $t = 0$, the correlation is equal to the colocated correlation parameter of the parsimonious Mat\'{e}rn cross-covariance function, i.e., $\rho$ with value set at 0.5. As $|t| \rightarrow \infty$, the colocated correlation decreases to zero. Moreover, as the multiple advections model gets farther and farther from the single advection model, the faster the drop of the colocated correlation to zero. This is what is shown by the plots in the first row (I). The colocated correlation declines more rapidly when $\mathbf{V}_{11}$ and $\mathbf{V}_{22}$ are negatively correlated compared to when they are positively correlated. The drop in the values of the colocated correlation is accelerated when $\mathbf{V}_{11}$ and $\mathbf{V}_{22}$ have different mean advections and this is what is displayed in the second (II) and third (III) rows of Figure~\ref{fig:fig2}. Overall, Figure~\ref{fig:fig2} conveys that failing to acknowledge multiple advections can lead to overestimation or underestimation of the true dependence between any two variables. In doing so, prediction accuracy may be affected.

In the models so far, the cross-advections $\mathbf{V}_{ij}$, $i \neq j$, cannot be prescribed independently of the temporal locations, i.e., the transport behavior experienced in the cross-covariance remains a function of the marginal advections $\mathbf{V}_{ii}$ and the temporal locations. An interesting problem is to find a form for $\mathbf{V}_{ij}$ such that 
\begin{equation} \label{eqn:proposed_single_advec}
\mathbf{C}(\mathbf{h}, u) =  \text{E}_{\boldsymbol{\mathcal{V}}} [ \{C_{ij}^S(\mathbf{h} - \mathbf{V}_{ij} u) \}_{i, j = 1}^{p} ],
\end{equation}
is valid. Here $\boldsymbol{\mathcal{V}}$ is the vector of marginal and cross-advections. Suppose from (\ref{eqn:proposed_single_advec}) that we build the covariance matrix $\boldsymbol{\Sigma}$ as follows:
\begin{equation} \label{eqn:full_covariance_matrix}
\boldsymbol{\Sigma} = \begin{bmatrix}
\mathbf{C}_{11, \mathbf{V}_{11}} & \mathbf{C}_{12, \mathbf{V}_{12}} & \cdots & \mathbf{C}_{1p, \mathbf{V}_{1p}} \\
\mathbf{C}_{21, \mathbf{V}_{21}} & \mathbf{C}_{22, \mathbf{V}_{22}} & \cdots & \mathbf{C}_{2p, \mathbf{V}_{2p}} \\
\vdots & \vdots & \ddots & \vdots \\
\mathbf{C}_{p1, \mathbf{V}_{p1}} & \mathbf{C}_{p2, \mathbf{V}_{p2}} & \cdots & \mathbf{C}_{pp, \mathbf{V}_{pp}} \\
\end{bmatrix} \in \mathbb{R}^{np \times np},
\end{equation}
where $\mathbf{C}_{ij, \mathbf{V}_{ij}} = \left( \text{E}_{\boldsymbol{\mathcal{V}}}\left[ C_{ij}^S \left\{ \mathbf{s}_l - \mathbf{s}_r - \mathbf{V}_{ij} (t_l - t_r) \right\} \right] \right)_{l, r = 1}^{n}  \in \mathbb{R}^{n \times n}$, for $i , j = 1, \ldots, p$. For the model in (\ref{eqn:proposed_single_advec}) to be valid, $\boldsymbol{\Sigma}$ has to be positive definite. By Theorem 1 in \cite{ip2015time}, $\boldsymbol{\Sigma}$ is positive definite if and only if the $(np \times np)$ matrix $\mathbf{K} $ with entries $\mathbf{K} = \left( \mathbf{C}_{ii, \mathbf{V}_{ii}}^{-1/2} \mathbf{C}_{ij, \mathbf{V}_{ij}} \mathbf{C}_{jj, \mathbf{V}_{jj}}^{-1/2} \right)_{i, j = 1}^{p}  \in \mathbb{R}^{np \times np}$ is positive definite. Here $\mathbf{A}^{1/2}$ is the symmetric square root of a square matrix $\mathbf{A}$ such that $\mathbf{A}^{1/2} \mathbf{A}^{1/2} = \mathbf{A}$. This result gives us more control regarding the transport or advection behavior in the cross-covariances. However, it remains a challenge to more precisely characterize $\mathbf{V}_{ij}$ such that $\mathbf{K}$ is indeed positive definite.

Manipulating the transport behavior in the cross-covariances can also be done by defining new dimensions in space or time and allowing variable specific advections in those extra dimensions. The following theorem establishes a Lagrangian spatio-temporal cross-covariance model that augments the spatial dimensions.

\begin{theo} \label{variable_specific_advec}
Let $\mathbf{V}_{11}, \mathbf{V}_{22}, \ldots, \mathbf{V}_{pp}$ be random vectors on $\mathbb{R}^d$ and $\mathbf{V}'_{11}, \mathbf{V}'_{22}, \ldots, \mathbf{V}'_{pp}$ be random vectors on $\mathbb{R}^{d'}$. If $C_{ij}^S(\mathbf{h}, \mathbf{h}')$ is a valid purely spatial stationary cross-covariance function on $\mathbb{R}^{d + d'}$, then
%Let $\mathbf{V}_{11}, \mathbf{V}_{22}, \ldots, \mathbf{V}_{pp}$ be random vectors on $\mathbb{R}^d$ and $\mathbf{V}'_{11}, \mathbf{V}'_{22}, \ldots, \mathbf{V}'_{pp}$ be random vectors on $\mathbb{R}^{d'}$. If $\mathbf{C}^{S}\{(\mathbf{h}, \mathbf{h}')\}$ is a valid purely spatial matrix-valued stationary covariance function on $\mathbb{R}^{d + d'}$, i.e., $\mathbf{C}^{S}(\mathbf{h}) = [C_{ij} \{(\mathbf{h}, \mathbf{h}')\} ]_{i,j=1}^{p},$ then
\begin{equation} \label{eqn:proposed_variable_specific_advec}
C_{ij}\{(\mathbf{h}, \mathbf{h}'_{ij}); t_1, t_2 \} = \normalfont \text{E}_{\boldsymbol{\tilde{\mathcal{V}}}} \{C_{ij}^S(\mathbf{h} - \mathbf{V}_{ii} t_1 +  \mathbf{V}_{jj} t_2, \mathbf{h}'_{ij} - \mathbf{V}'_{ii} t_1 + \mathbf{V}'_{jj} t_2 )\},
\end{equation}
where $\mathbf{h}'_{ij} = \mathbf{s}'_{ii} - \mathbf{s}'_{jj}$, for $\mathbf{s}'_{ii}, \ldots, \mathbf{s}'_{pp} \in \mathbb{R}^{d'}$, and the expectation is taken with respect to the joint distribution of $\boldsymbol{\tilde{\mathcal{V}}} = \{ (\mathbf{V}_{11}^{\top}, \mathbf{V}_{11}^{'\top}), (\mathbf{V}_{22}^{\top}, \mathbf{V}_{22}^{'\top}), \ldots, (\mathbf{V}_{pp}^{\top}, \mathbf{V}_{pp}^{'\top})\}^{\top}$, is a valid spatio-temporal cross-covariance function on $\mathbb{R}^{d + d'} \times \mathbb{R}$ provided that the expectation exists.
\end{theo}
When $d' = 1$, $s'_{ii} \in \mathbb{R}$ can possibly be the altitude or the location in the $z$-axis at which variable $i$ was taken and $\text{V}'_{ii} \in \mathbb{R}$ is the component of the advection velocity along that axis. Augmenting the temporal dimensions can also be done similarly. However, introducing a vector of temporal locations brings an unnatural physical interpretation to the Lagrangian transport phenomenon. Even in the classical univariate model of \cite{cox1988} in (\ref{eqn:coxishameq}), the form of the Lagrangian shift when the temporal argument becomes a vector is nontrivial and has not yet been explored anywhere. A Lagrangian shift $\mathbf{h} - \ddot{\mathbf{V}} \mathbf{u}$, where $\mathbf{h} \in \mathbb{R}^d$, $\mathbf{u} \in \mathbb{R}^{d''}$, and $\ddot{\mathbf{V}} \in \mathbb{R}^{d \times d''}$ is the advection velocity matrix, can be pursued when faced with multiple dimensions in time. The columns of $\ddot{\mathbf{V}}$ indicate the component of the transport velocity in every dimension of time. However, due to a lack of useful physical interpretation of Lagrangian models with an advection velocity matrix, we discuss only~(\ref{eqn:proposed_variable_specific_advec}) and pursue the idea of advection velocity matrix in another work.

We simulate from (\ref{eqn:proposed_variable_specific_advec}) using the closed forms in (\ref{eqn:proposed-nonfrozen-multi-advec-marginal}) and (\ref{eqn:proposed-nonfrozen-multi-advec-cross}) such that $Z_1$ is taken at $(\mathbf{s}^{\top}, 0.2)^{\top}$ and $Z_2$ at $(\mathbf{s}^{\top}, 0)^{\top}$, i.e., $Z_1$ and $Z_2$ have the same locations in the $xy$-axis but are separated 0.2 units away in the $z$-axis. Moreover, we set the advection of $Z_1$ and $Z_2$ in the $xy$-axis to be the same, i.e., $\text{E}(\mathbf{V}_{11}) = \text{E}(\mathbf{V}_{22})  = (0.1, 0.1)^{\top}$, but we augment them with different vertical components. In particular, we set $\text{V}'_{11}$ and $\text{V}'_{22}$ such that $Z_1$ gets transported downwards, while $Z_2$ gets transported upwards, i.e., $\text{E}(\text{V}'_{11}) = -0.05$ and $\text{E}(\text{V}'_{22}) = 0.05$. Furthermore, we consider three different strengths of dependence between the vertical components, namely (a) $\text{cov}(\text{V}'_{11}, \text{V}'_{22}) = 0.9$, (b) $\text{cov}(\text{V}'_{11}, \text{V}'_{22}) = 0$, and (c) $\text{cov}(\text{V}'_{11}, \text{V}'_{22}) = -0.9$. The realizations of the resulting bivariate Lagrangian spatio-temporal random field are shown in Panels (d)-(f) in Figure~\ref{fig:fig1}. %From the figures, it can be seen that more sophisticated spatio-temporal dependencies spring from adding dimensions in space. For instance, in Figure~\ref{fig:fig2}(d), the maximum cross-covariance occurs at $t_1 = t_2 = 2$ and not at $t_1 = t_2 = 1$, which is the case in Figure~\ref{fig:fig2}(a). This is expected because while $Z_1$ and $Z_2$ are colocated in $\mathbb{R}^2$, they are actually 0.2 units apart in $\mathbb{R}^3$. As $Z_1$ travels downwards and $Z_2$ travels upwards, they decrease the spatial separation between them, which increases their spatial dependence. However, this behavior is not manifested in Figure~\ref{fig:fig2}(e)-(f). This is because independent or negatively highly correlated advections bring about greater reduction in the maximum purely spatial cross-covariance. That is, $Z_1$ and $Z_2$ may be colocated in $\mathbb{R}^3$ after some time but variability from the mean advections distorts the transported purely spatial random fields, thereby reducing the maximum purely spatial cross-covariance attainable.

Another way to introduce multiple advections in the Lagrangian framework while remaining stationary in time is by using latent uncorrelated univariate transported purely spatial random fields, each influenced by different advection velocities. \cite{salvana2020nonstationary} suggested such models with latent transported purely spatial random fields that are second-order nonstationary. Their proposed model of course remains valid when the latent transported purely spatial random fields are second-order stationary. We formalize such models in the following theorem. 

\begin{theo}\label{lmc_multiple}
Let $\mathbf{V}_{r},\;r=1,\ldots,R$, be random vectors on $\mathbb{R}^d$. If $\rho_{r}(\mathbf{h})$ are valid univariate stationary correlation functions on $\mathbb{R}^d$, then
\begin{equation}\label{eqn:spacetimelmc_expectation}
\mathbf{C}(\mathbf{h}, u) = \sum_{r=1}^{R} \normalfont \text{E}_{\mathbf{V}_r}\left\{\rho_r(\mathbf{h} - \mathbf{V}_r u )\right\} \mathbf{T}_r
\end{equation}
is a valid spatio-temporal matrix-valued stationary cross-covariance function on $\mathbb{R}^d \times \mathbb{R}$, for any $1\leq R\leq p$ and $\mathbf{T}_r$, $r = 1, \ldots, R$, are positive semi-definite matrices.
\end{theo}

The model in (\ref{eqn:spacetimelmc_expectation}) is the resulting Lagrangian spatio-temporal cross-covariance function of the following multivariate spatio-temporal process: 
\begin{equation}\label{eqn:lmcprocess}
\mathbf{Z}(\mathbf{s}, t) = \mathbf{A} \mathbf{W}(\mathbf{s}, t) = \mathbf{A} [ W_1 (\mathbf{s} - \mathbf{V}_1 t), W_2 (\mathbf{s} - \mathbf{V}_2 t), \ldots, W_R (\mathbf{s} - \mathbf{V}_R t) ] ^{\top},
\end{equation}
where $\mathbf{A}$ is a $p\times R$ matrix and the components of $\mathbf{W}(\mathbf{s}, t) \in \mathbb{R}^R$ are independent but not identically distributed. Each component $W_r$ has a univariate Lagrangian spatio-temporal stationary correlation function $\rho_r (\mathbf{h} - \mathbf{V}_r u),\;r=1,\ldots,R.$ Here, $\mathbf{T}_r = \mathbf{a}_r \mathbf{a}_r^{\top}$, where $\mathbf{a}_r$ is the $r$th column of $\mathbf{A}$. Moreover, when $\mathbf{V}_{1} = \mathbf{V}_{2} = \cdots = \mathbf{V}_{R} = \mathbf{V}$, we return to the single advection velocity vector case and retrieve the Lagrangian spatio-temporal version of the linear model of coregionalization (LMC); see \cite{gelfand2002multivariate} and \cite{wackernagel1998} for the discussion of such class of purely spatial cross-covariance functions.

%%%%%%%%%%%%%%%%%%%%%%%%%%%%%%%%%%%%%%%%%%%%%%%%

\section{Estimation} \label{sec:estimation}

In this section, we outline a viable estimation procedure involving Lagrangian spatio-temporal cross-covariance functions with multiple advections. Suppose $\mathbf{Y} = \{ \mathbf{Y}(\mathbf{s}_1, t_1)^{\top}, \ldots, \mathbf{Y}(\mathbf{s}_n, t_n)^{\top} \}^{\top} \in \mathbb{R}^{np}$ is an $np$-vector of multivariate spatio-temporal observations such that $n$ is the total number of spatio-temporal locations and $p$ is the number of variables. Assume that the mean function in~(\ref{eqn:linear_model}) can be characterized as a linear combination of some covariates $X_1, X_2, \ldots, X_M$. Denote by $\boldsymbol{\beta} = (\boldsymbol{\beta}_1^{\top}, \ldots, \boldsymbol{\beta}_p^{\top})^{\top} \in \mathbb{R}^{Mp}$ the vector of mean parameters, where $\boldsymbol{\beta}_i = (\beta_{1, i}, \ldots, \beta_{M, i})^{\top} \in \mathbb{R}^{M}$, for $i = 1, \ldots, p$, and $\mathbf{X} = \left\{ \mathbf{I}_p \otimes  \mathbf{X}(\mathbf{s}_1, t_1)^{\top}, \mathbf{I}_p \otimes \mathbf{X}(\mathbf{s}_2, t_2)^{\top}, \ldots, \mathbf{I}_p \otimes \mathbf{X}(\mathbf{s}_n, t_n)^{\top} \right\}^{\top} \in \mathbb{R}^{np\times  Mp}$, where $\mathbf{X}(\mathbf{s}, t) = \left\{X_1(\mathbf{s}, t), \ldots, X_M(\mathbf{s}, t) \right\}^{\top}  \in \mathbb{R}^{M}$. The model in (\ref{eqn:linear_model}) becomes $\mathbf{Y}(\mathbf{s}, t) = \left\{  \mathbf{I}_p \otimes \mathbf{X}(\mathbf{s}, t)^{\top} \right\} \boldsymbol{\beta} + \mathbf{Z}(\mathbf{s},t). $ Furthermore, denote by $\boldsymbol{\Sigma}(\boldsymbol{\Theta})$ the $n p \times n p$ covariance matrix, parameterized by $\boldsymbol{\Theta} \in \mathbb{R}^{q}$ such that $\boldsymbol{\Sigma}(\boldsymbol{\Theta}) = [ \left\{ C_{ij} (\mathbf{s}_l - \mathbf{s}_r; t_l, t_r | \boldsymbol{\Theta}) \right\}_{i,j = 1}^{p} ]_{l, r = 1}^{n}$. The mean parameters, $\boldsymbol{\beta}$, and the cross-covariance parameters, $\boldsymbol{\Theta}$, are estimated via restricted maximum likelihood (REML) estimation which proceeds by maximizing, through an iterative procedure \citep[Equations 4 and 5]{cressie1996asymptotics, shor2019estimating}:
\begin{eqnarray}\label{eqn:loglikelihood_general}
l_{\text{REML}}(\boldsymbol{\Theta}, \boldsymbol{\beta}; \mathbf{Y}) & = & l_{\text{MLE}}(\boldsymbol{\Theta}, \boldsymbol{\beta}; \mathbf{Y}) + \frac{M p}{2} \log(2 \pi) + \frac{1}{2} \log | \mathbf{X}^{\top}  \mathbf{X} |  - \frac{1}{2} \log | \mathbf{X}^{\top} \boldsymbol{\Sigma}(\boldsymbol{\Theta})^{-1} \mathbf{X} |, \label{eqn:loglikelihood_reml} \\
l_{\text{MLE}}(\boldsymbol{\Theta}, \boldsymbol{\beta}; \mathbf{Y}) & = & - \frac{n p}{2} \log(2 \pi) - \frac{1}{2} \log | \boldsymbol{\Sigma}(\boldsymbol{\Theta}) |  - \frac{1}{2} (\mathbf{Y} - \mathbf{X} \boldsymbol{\beta} )^{\top} \boldsymbol{\Sigma}(\boldsymbol{\Theta})^{-1} (\mathbf{Y} - \mathbf{X} \boldsymbol{\beta} ), \label{eqn:loglikelihood_mle}  \hspace*{1cm}
\end{eqnarray}
where $l_{\text{MLE}}$ denotes the usual log-likelihood under maximum likelihood estimation (MLE). The iteration procedure begins with an initialization of $\boldsymbol{\beta}$ which we set to the ordinary least squares (OLS) estimate, $\hat{\boldsymbol{\beta}}_{\text{OLS}} = ( \mathbf{X}^{\top} \mathbf{X} )^{-1} \mathbf{X}^{\top} \mathbf{Y}$, and which we plug-in to the log-likelihood equations above wherever $\boldsymbol{\beta}$ appears. 

Next, we estimate $\boldsymbol{\Theta}$ in a multi-step fashion. Splitting the estimation problem into several parts has been routinely employed when groups of parameters in the cross-covariance function can be estimated sequentially \citep{apanasovich2010cross, bourotte2016flexible, qadir2020flexible}. Furthermore, it has been established that under some fairly general conditions, the multi-step procedure yields consistent estimators of the parameters in the last step \citep{murphy2002estimation, zhelonkin2012robustness, greene2014econometric}. The parameters of any Lagrangian spatio-temporal cross-covariance function with multiple advections such that $\boldsymbol{\mathcal{V}} \sim \mathcal{N}_{pd} (\boldsymbol{\mu}_{\boldsymbol{\mathcal{V}}}, \boldsymbol{\Sigma}_{\boldsymbol{\mathcal{V}}} )$ can be partitioned in two, i.e., $\boldsymbol{\Theta} = \left( \boldsymbol{\theta}_{-\boldsymbol{\Sigma}_{\boldsymbol{\mathcal{V}}}}^{\top}, \boldsymbol{\theta}_{\boldsymbol{\Sigma}_{\boldsymbol{\mathcal{V}}}}^{\top} \right)^{\top} \in \mathbb{R}^q$, where $\boldsymbol{\theta}_{-\boldsymbol{\Sigma}_{\boldsymbol{\mathcal{V}}}}$ is the vector of parameters excluding the parameters associated with $\boldsymbol{\Sigma}_{\boldsymbol{\mathcal{V}}}$ and $\boldsymbol{\theta}_{\boldsymbol{\Sigma}_{\boldsymbol{\mathcal{V}}}}$ are the parameters that build $\boldsymbol{\Sigma}_{\boldsymbol{\mathcal{V}}}$. Note that $\boldsymbol{\theta}_{-\boldsymbol{\Sigma}_{\boldsymbol{\mathcal{V}}}}$ includes all purely spatial parameters and $\boldsymbol{\mu}_{\boldsymbol{\mathcal{V}}}$. We found that the parameters associated with $\boldsymbol{\mu}_{\boldsymbol{\mathcal{V}}}$ should not be estimated alongside those of $\boldsymbol{\Sigma}_{\boldsymbol{\mathcal{V}}}$, otherwise, the MLE or REML estimates will converge to a local maximum and we might obtain estimates that are far from the true value. The elements of $\boldsymbol{\Theta}$ are estimated sequentially as follows:
\begin{enumerate}[Step 1:]

\item Initialize $\boldsymbol{\Sigma}_{\boldsymbol{\mathcal{V}}}$ and run the optimization through the candidate vectors $\boldsymbol{\theta}_{-\boldsymbol{\Sigma}_{\boldsymbol{\mathcal{V}}}}$ and find $\boldsymbol{\hat{\theta}}_{-\boldsymbol{\Sigma}_{\boldsymbol{\mathcal{V}}}}$ that maximizes (\ref{eqn:loglikelihood_general}). This is likened to fitting a frozen field multiple advections model.

\item Fit the non-frozen field version of the frozen multiple advections model in Step 1 by finding $\boldsymbol{\hat{\theta}}_{\boldsymbol{\Sigma}_{\boldsymbol{\mathcal{V}}}}$ that maximizes (\ref{eqn:loglikelihood_general}) while fixing the other parameters to $\boldsymbol{\hat{\theta}}_{-\boldsymbol{\Sigma}_{\boldsymbol{\mathcal{V}}}}$. To ensure $\boldsymbol{\Sigma}_{\boldsymbol{\mathcal{V}}}$ remains positive definite, its entries are parameterized via its Cholesky decomposition. 

\end{enumerate}

% %https://people.csail.mit.edu/xiuming/docs/tutorials/reml.pdf

Once $\hat{\boldsymbol{\Theta}} = \left( \boldsymbol{\hat{\theta}}_{-\boldsymbol{\Sigma}_{\boldsymbol{\mathcal{V}}}}^{\top}, \boldsymbol{\hat{\theta}}_{\boldsymbol{\Sigma}_{\boldsymbol{\mathcal{V}}}}^{\top} \right)^{\top}$ is obtained, we solve for $\hat{\boldsymbol{\beta}}_{\text{GLS}}$, where $\hat{\boldsymbol{\beta}}_{\text{GLS}}$ is the vector of estimates of the regression coefficients via generalized least squares (GLS) of the form $\hat{\boldsymbol{\beta}}_{\text{GLS}} = \{ \mathbf{X}^{\top} \boldsymbol{\Sigma}(\hat{\boldsymbol{\Theta}})^{-1} \mathbf{X} \}^{-1} \mathbf{X}^{\top} \boldsymbol{\Sigma}(\hat{\boldsymbol{\Theta}})^{-1} \mathbf{Y}$ and loop again through the above multi-step estimation of $\boldsymbol{\Theta}$. The procedure is terminated when a stopping criterion is reached.

 \begin{figure}[t!]
 \centering
	\includegraphics[scale=0.35]{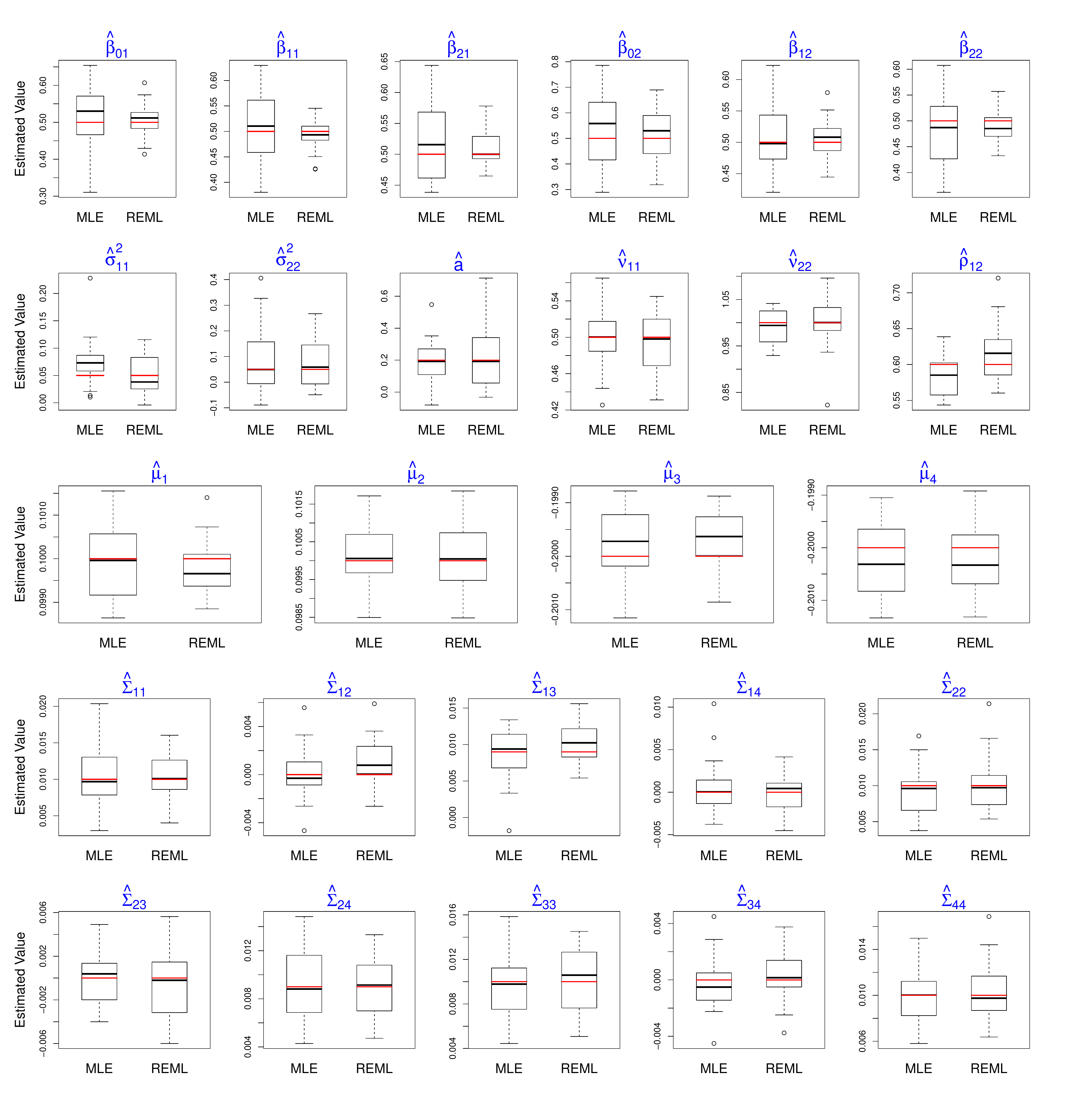}
	\caption{Boxplots of the parameter estimates using MLE and REML. The true parameter values are highlighted in red.}
    \label{fig:mle_vs_reml}
\end{figure}

We simulate 100 realizations of Gaussian bivariate spatio-temporal random fields of the form (\ref{eqn:linear_model}) with $\mu_i (\mathbf{s}, t)  = \mu_i (\mathbf{s}) = 0.5 + 0.5 s_x + 0.5 s_y$, $i = 1, 2$, and $\mathbf{Z}(\mathbf{s}, t)$ has a cross-covariance function given in Theorem~\ref{theorem_multiple_advec_nonfrozen}, with purely spatial parameters as in Figure~\ref{fig:fig1} and $\rho = 0.6$, on a $20 \times 20$ grid in the unit square, at time $t = 0, 1, \ldots, 4$. We set $\text{E}(\mathbf{V}_{11}) = (0.1, 0.1)^{\top}$, $\text{E}(\mathbf{V}_{22}) = (-0.2, -0.2)^{\top}$, and $\mathbf{V}_{11}$ and $\mathbf{V}_{22}$ to be independent random vectors. Figure~\ref{fig:mle_vs_reml} plots the estimates obtained using the proposed multi-step estimation under MLE~(\ref{eqn:loglikelihood_mle}) and REML~(\ref{eqn:loglikelihood_reml}). Note that we denote the entries of $\boldsymbol{\mu}_{\boldsymbol{\mathcal{V}}}$ and $\boldsymbol{\Sigma}_{\boldsymbol{\mathcal{V}}}$ as $\boldsymbol{\mu}_{\boldsymbol{\mathcal{V}}} = (\mu_1, \mu_2, \mu_3, \mu_4)^{\top}$ and $\boldsymbol{\Sigma}_{\boldsymbol{\mathcal{V}}} = (\Sigma_{ij})_{i, j = 1}^{4}$, respectively. The boxplots show that despite having 26 parameters to estimate, the proposed estimation procedure was able to obtain parameter estimates that are satisfactorily close to the true values both under MLE and REML. However, the variances of the parameter estimates, especially those associated with the mean function, tend to be smaller under REML. Hence, we employ REML in the remainder of this paper.
 
When the observations at $N$ spatial locations are obtained at regular intervals and there is negligible dependence between observations that are very distant in the temporal sense, according to \cite{stein2005statistical}, for large number of temporal locations, $T$, (\ref{eqn:loglikelihood_general}) can be approximated as:
\begin{equation}\label{loglikelihood_approximation}
l_{\text{REML}}(\boldsymbol{\Theta}, \boldsymbol{\beta}; \mathbf{Y}) \approx  l_{\text{REML}}(\boldsymbol{\Theta}, \boldsymbol{\beta}; \mathbf{Y}_{1, t^*}) + \sum_{j = t^* + 1}^T l_{\text{REML}}(\boldsymbol{\Theta}, \boldsymbol{\beta}; \mathbf{Y}_{j} | \mathbf{Y}_{j - t^*, j - 1}),
\end{equation}
where $\mathbf{Y}_{a, b} = ( \mathbf{Y}_a^{\top}, \ldots, \mathbf{Y}_b^{\top})^{\top} \in \mathbb{R}^{N p t^*}$, ${\mathbf{Y}_{t}} = \{\mathbf{Y} (\mathbf{s}_1, t)^{\top}, \ldots, \mathbf{Y} (\mathbf{s}_N, t)^{\top} \}^{\top} \in \mathbb{R}^{Np}$, for $a < b$, and $t^*$ specifies the number of consecutive temporal locations included in the conditional distribution. Here $l_{\text{REML}}(\boldsymbol{\Theta}, \boldsymbol{\beta}; \mathbf{Y}_{j} | \mathbf{Y}_{j - t^*, j - 1})$ is the log-likelihood function based only on the vector of spatio-temporal measurements $\mathbf{Y}_{j - t^*, j-1} = \big( \mathbf{Y}_{j - t^*}^{\top}, \ldots, \mathbf{Y}_{j - 1}^{\top} \big)^{\top}$.

\section{Simulation Study} \label{sec:simulation}

Under the single advection setting and for $p = 1$, \cite{salvana2021lagrangian} showed that when the components of $\mathbf{V} \sim \mathcal{N}_d (\boldsymbol{\mu}_{\mathbf{V}}, \boldsymbol{\Sigma}_{\mathbf{V}})$ have zero-mean, the same variance, and are uncorrelated, i.e., $\boldsymbol{\mu}_{\mathbf{V}} = \mathbf{0}$ and $\boldsymbol{\Sigma}_{\mathbf{V}} = \sigma_{\mathbf{V}}^2 \mathbf{I}_d$, for any $\sigma_{\mathbf{V}}^2 > 0$, the univariate Lagrangian spatio-temporal models with normal scale-mixture $C^S$ reduce to univariate non-Lagrangian spatio-temporal isotropic covariance functions belonging to the Gneiting class \citep{gneiting2002}. A breakdown on any of the above-mentioned restrictions dichotomizes univariate Lagrangian spatio-temporal models from their non-Lagrangian counterparts. Their simulation studies can be adopted for $p > 1$ and similar conclusions can be drawn.

When faced with a multivariate Lagrangian spatio-temporal random field with multiple advections, one can either fit multivariate Lagrangian spatio-temporal models with multiple advections, as in (\ref{eqn:locally_stationary}), or marginally fit on each variable a univariate Lagrangian spatio-temporal model, as in (\ref{eqn:coxishameq}). While it has been shown in the literature that multivariate modeling generally yields lower prediction errors as the presence of the other variables essentially increases the sample size of one variable \citep{genton2015cross, zhang2015doesn, salvana2021high}, it remains to be explored how the dependence between any two advection velocities affects prediction accuracy. To answer this inquiry, we perform experiments to identify scenarios where multivariate Lagrangian spatio-temporal models with multiple advections are favorable over multiple univariate Lagrangian spatio-temporal ones. Another objective of this section is to show the consequences of using a bivariate Lagrangian spatio-temporal cross-covariance function with single advection to model a bivariate Lagrangian spatio-temporal random field with advections $\mathbf{V}_{11} \sim \mathcal{N}_d \{\text{E}(\mathbf{V}_{11}), \text{var}(\mathbf{V}_{11}) \}$ and $\mathbf{V}_{22} \sim \mathcal{N}_d \{\text{E}(\mathbf{V}_{22}), \text{var}(\mathbf{V}_{22})\}$ such that $\text{E}(\mathbf{V}_{11}) = \text{E}(\mathbf{V}_{22})$ and $\text{var}(\mathbf{V}_{11}) = \text{var}(\mathbf{V}_{22})$ but $\mathbf{V}_{11} \neq \mathbf{V}_{22}$. Such bivariate random fields appear to be driven by a single advection when in fact they are not. The nature of dependence between $\mathbf{V}_{11}$ and $\mathbf{V}_{22}$ may or may not be useful in modeling or prediction. In this section, we aim to expose such consequences.

%Here we provide extensive simulation studies that will show the impact of such misspecification to the parameter estimates and the loss of efficiency in the predictions. Section 2 already illustrated some distinctions between the two by a direct comparison of their simulated realizations and cross-covariance function values. 
 
%\section{Non-Frozen vs. Frozen Field Models}\label{sec:simulation}
\subsection{Design}

%\begin{figure}[h!]
% \centering
%\includegraphics[scale=0.47]{fig8.pdf}
%  \caption{\small Rows 1-3: Empirical correlations and cross-correlation at $u=1$ for M4 simulations with different $\Lambda$. Row 4: Averaged estimated distribution of the advection velocity vector under M1 with different $\Lambda$.}
%    \label{fig:m4data}
%\end{figure}

All the simulation studies are framed under the assumption that $d = 2$ and $p = 2$. Consider the following Lagrangian spatio-temporal models:
\begin{compactitem}
\item M1: univariate Lagrangian spatio-temporal model in (\ref{eqn:schlather_nonfrozen}), where $C^S$ is the Mat\'{e}rn covariance function with purely spatial parameters $\sigma, a, \nu$;

\item M2: bivariate Lagrangian spatio-temporal model with single advection, i.e.,
\begin{equation*} \label{eqn:single_advec}
C_{ij}(\mathbf{h}, u) = \frac{ \rho \sigma_{ii}\sigma_{jj} }{\sqrt{|\mathbf{I}_d + \boldsymbol{\Sigma}_{\mathbf{V}}u^2|}} \mathcal{M} \{ \left(\mathbf{h} - \boldsymbol{\mu}_{\mathbf{V}}u\right)^{\top} \left(\mathbf{I}_d + \boldsymbol{\Sigma}_{\mathbf{V}} u^2\right)^{-1} \left(\mathbf{h}- \boldsymbol{\mu}_{\mathbf{V}}u\right); a, \nu_{ij} \}, \quad i, j = 1, 2,
\end{equation*}
where $\mathcal{M}(\|\mathbf{h}\|; a, \nu)$ is the univariate Mat\'{e}rn correlation with spatial range and smoothness parameters $a$ and $\nu$, respectively; and

\item M3: bivariate Lagrangian spatio-temporal model with multiple advections in (\ref{eqn:proposed-nonfrozen-multi-advec-marginal}) and (\ref{eqn:proposed-nonfrozen-multi-advec-cross}), where $C_{ij}^S$ is the parsimonious Mat\'{e}rn cross-covariance function, with purely spatial parameters $\rho, \sigma_{ij}, a, \nu_{ij}$, $i, j = 1, 2$.
\end{compactitem}
We simulate 100 realizations of zero-mean bivariate Gaussian spatio-temporal random fields, $Z_1 (\mathbf{s}, t) $ and $Z_2(\mathbf{s}, t)$, with purely spatial parameters as in Figure~\ref{fig:fig1}, containing $N = 529$ spatial observations, on a $23 \times 23$ grid in the unit square, at time $t = 0, 1, \ldots, 5$. This number of spatio-temporal locations is chosen to mimic the setup in the real data application in Section~\ref{sec:application}. In the following experiments, we remove the observations at $t = 5$ and use the remaining observations to fit the models. Upon obtaining the parameter estimates, we predict the previously removed values using simple kriging for the univariate model and simple cokriging for the bivariate models and report the error of the predictions measured by the Mean Square Error (MSE), $\text{MSE} = \frac{1}{p N} \sum_{i = 1}^p \sum_{l = 1}^{N} \{ \hat{Z}_{i}(\mathbf{s}_{l}, 5) - Z_{i}(\mathbf{s}_{l}, 5) \}^2$. 

For the first experiment, we try to understand the effect of the dependence between the two advection velocities, $\mathbf{V}_{11}$ and $\mathbf{V}_{22}$, on the accuracy of predictions. We simulate from M3 under different assumptions on the joint distribution of $\mathbf{V}_{11}$ and $\mathbf{V}_{22}$, namely, $\text{E}(\mathbf{V}_{11}) = (0.1, 0.1)^{\top}$, $\text{E}(\mathbf{V}_{22}) = (-0.1, 0.1)^{\top}$ and (a) $\mathbf{V}_{11} = 0.9 \mathbf{V}_{22}$, (b) $\mathbf{V}_{11}$ and $\mathbf{V}_{22}$ are independent, and (c) $\mathbf{V}_{11} = - 0.9 \mathbf{V}_{22}$, for several values of the spatial cross-correlation parameter, i.e., $\rho = \pm 0.3, \pm 0.6, \pm 0.9$. On the simulated values, we fit the true model, M3, and a simpler alternative model, M1. For the second experiment, we try to reveal the consequences of using a single advection model instead of a multiple advections model. We simulate from M3 such that $\text{E}(\mathbf{V}_{11})  = \text{E}(\mathbf{V}_{22}) = (0.1, 0.1)^{\top}$ with different $\text{cov}(\mathbf{V}_{11}, \mathbf{V}_{22})$, namely, (d) $\text{cov}(\mathbf{V}_{11}, \mathbf{V}_{22}) = 0.001 \left(\protect\begin{smallmatrix} 1 & 0.9  \\
0.9 & 1 \protect\end{smallmatrix}\right) \otimes \mathbf{I}_2 $, (e) $\text{cov}(\mathbf{V}_{11}, \mathbf{V}_{22}) = 0.001  \mathbf{I}_4 $, (f) $\text{cov}(\mathbf{V}_{11}, \mathbf{V}_{22}) = 0.001 \left(\protect\begin{smallmatrix} 1 & -0.9  \\
-0.9 & 1 \protect\end{smallmatrix}\right) \otimes \mathbf{I}_2$, and $\rho = \pm 0.3, \pm 0.6, \pm 0.9$. Scenarios (d) and (f) represent the highly positive and negative dependence between the corresponding components of $\mathbf{V}_{11}$ and $\mathbf{V}_{22}$, respectively, while (e) establishes that $\mathbf{V}_{11}$ and $\mathbf{V}_{22}$ are independent. Both the true model, M3, and a simpler model, M2, are fitted to the simulated random fields from M3. 

%$\hat{\mathbf{Z}}_{0} = \mathbf{c}^{\top} \boldsymbol{\Sigma}(\hat{\boldsymbol{\Theta}})^{-1} \mathbf{Z}_{obs},$ where $\mathbf{c} =  \left[ \left\{ C_{ij} (\mathbf{s}_{obs_l} - \mathbf{s}_{obs_r}, t_{obs_l}, t_{obs_r}; \hat{\boldsymbol{\Theta}}) \right\}_{i,j = 1}^{2} \right]_{l, r = 1}^{n_{obs}}$, $\mathbf{Z}_{obs} = \{ \mathbf{Z} (\mathbf{s}_{obs_1}, t_{obs_1})^{\top}, \ldots, \mathbf{Z}(\mathbf{s}_{obs_{n_{obs}}}, t_{obs_{n_{obs}}})^{\top} \}^{\top}$, and $\mathbf{Z} (\mathbf{s}_{obs}, t_{obs}) = \{Z_1 (\mathbf{s}_{obs}, t_{obs}), Z_2 (\mathbf{s}_{obs}, t_{obs})\}^{\top} $, such that $n_{obs} = N_{obs} T$, 

%An objective of this study is to examine the effect of the degree of dependence between $\mathbf{V}_1$ and $\mathbf{V}_2$ to out-of-sample spatio-temporal prediction accuracy. A second objective is to quantify the loss of efficiency when a simpler model, the Lagrangian spatio-temporal model with single advection velocity vector, is used given that the true model contains multiple advection velocity vectors. 

%For the first objective, we focus our attention on the cross-covariance between $\mathbf{V}_1$ and $\mathbf{V}_2$ and consider different parameter combinations characterizing the joint normal distribution of $\mathbf{V}_1$ and $\mathbf{V}_2$ for the model in (\ref{eqn:proposed-nonfrozen-multi-advec}) listed as follows.

\subsection{Results and Analysis}

 \begin{figure}[t!]
 \centering
	\includegraphics[scale=0.45]{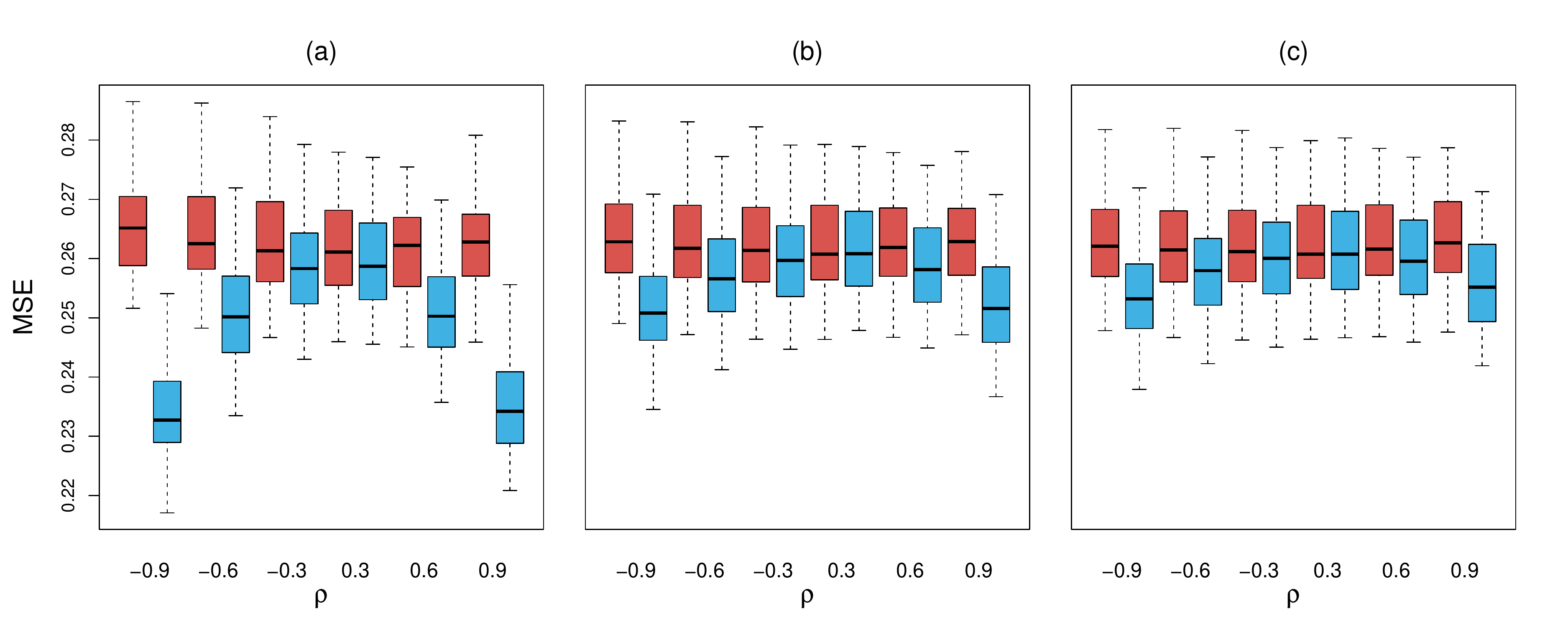}
	\caption{Boxplots of the MSEs under different assumptions on the joint distribution of $\mathbf{V}_{11}$ and $\mathbf{V}_{22}$, namely, (a) $\mathbf{V}_{11} = 0.9 \mathbf{V}_{22}$, (b) $\mathbf{V}_{11}$ and $\mathbf{V}_{22}$ are independent, and (c) $\mathbf{V}_{11} = - 0.9 \mathbf{V}_{22}$, when M1 (red) and M3 (blue) are fitted to data generated from M3 with different values of $\rho$.}
    \label{fig:simulation_study1}
\end{figure}

In the first experiment, we compare the accuracy of the predictions between the two modeling paradigms, namely, marginally fitting multiple univariate Lagrangian spatio-temporal models with different advections (M1) and fitting a bivariate Lagrangian spatio-temporal model with multiple advections (M3), given that the true model is the latter. Figure~\ref{fig:simulation_study1} summarizes the boxplots of the MSEs under different strengths of dependence between $\mathbf{V}_{11}$ and $\mathbf{V}_{22}$ and different values of the colocated correlation $\rho$. It can be seen that the prediction performance of M1 is fairly similar regardless of the true value of $\rho$ and the dependence between the different advections. Using a bivariate model (M3), on the other hand, improves predictions for nonzero $\rho$, with higher gains in accuracy as $\rho$ gets farther away from 0. Furthermore, M3 yields more accurate predictions when the true model consists of a nonzero $\rho$ with advection vectors that are highly dependent. When utilizing M1 over M3, one disregards a possible purely spatial variable dependence and possible dependence between the advections, which can help improve predictions. The reductions in the prediction errors are different and they depend on how correlated $\mathbf{V}_{11}$ and $\mathbf{V}_{22}$ are. When a random field is generated with a high $|\rho|$, combined with positively dependent $\mathbf{V}_{11}$ and $\mathbf{V}_{22}$, the prediction performance between M1 and M3 are strikingly different. As the advection velocities become independent or negatively dependent, the difference is not that huge. This is because independent or negatively dependent $\mathbf{V}_{11}$ and $\mathbf{V}_{22}$ dampen the colocated correlation between $Z_1$ and $Z_2$ (see Section 2, Figure~\ref{fig:fig2}) which substantially reduces the information that can be borrowed from $Z_1$ to help predict $Z_2$, or vice versa. When this happens, separately modeling and predicting $Z_1$ and $Z_2$ becomes as good as jointly modeling and predicting them.

 \begin{figure}[t!]
 \centering
	\includegraphics[scale=0.34]{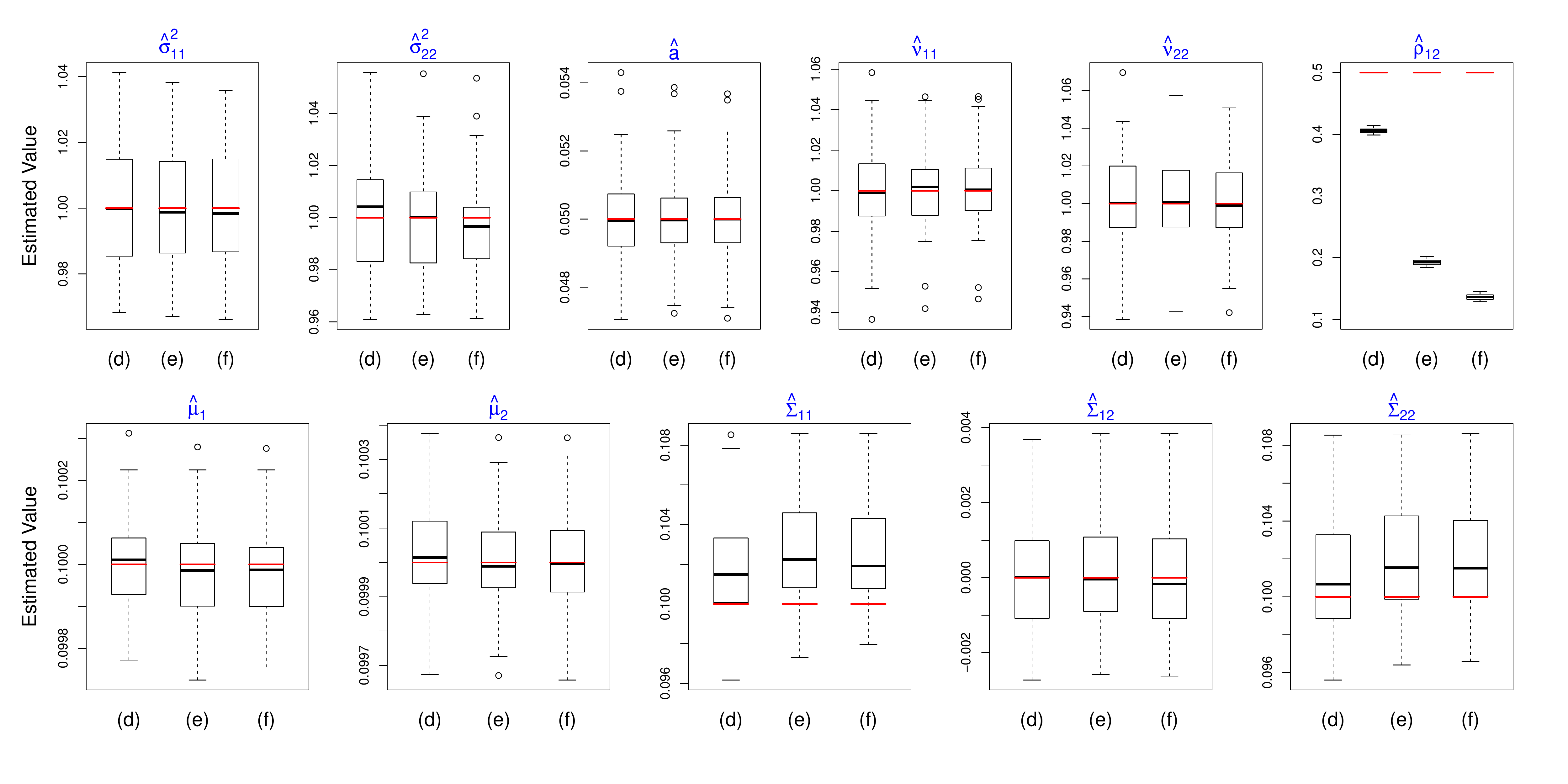}
	\caption{Boxplots of the parameter estimates of M2 under REML when it is fitted to data generated from M3 under scenarios (d), (e), and (f).  The true parameter values are highlighted in red.}
    \label{fig:simulation_study2}
\end{figure}

In the second experiment, we study the effect of fitting a single advection model to a bivariate random field that appears to be simulated from a single advection model on two fronts: 1) the estimates of the common parameters; and 2) the prediction performances of models M2 and M3. Figure~\ref{fig:simulation_study2} gives the summary of the parameter estimates of M2 under REML. Clearly, the most impacted parameter is $\rho$. As the dependence between $\mathbf{V}_{11}$ and $\mathbf{V}_{22}$ goes from highly positive to highly negative, $\hat{\rho} \rightarrow 0$. This is expected for the following reason. The cross-covariance function in M2 is stationary in space and time, which implies that the purely spatial colocated correlation is constant. M3, on the other hand, has a cross-covariance function that is nonstationary in time with a time-varying and decreasing purely spatial colocated correlation as $t$, i.e., $t = t_1 = t_2$, moves away from 0. Furthermore, the more negatively dependent $\mathbf{V}_{11}$ and $\mathbf{V}_{22}$ are, the faster the decline in the colocated correlation; see Section 2, Figure~\ref{fig:fig2}. Thus, when a model that can only handle a constant purely spatial colocated correlation parameter is fitted to a bivariate random field that possesses decreasing purely spatial colocated correlation, the optimization routine is required to find the best compromise for $\hat{\rho}$.

 \begin{figure}[t!]
 \centering
	\includegraphics[scale=0.45]{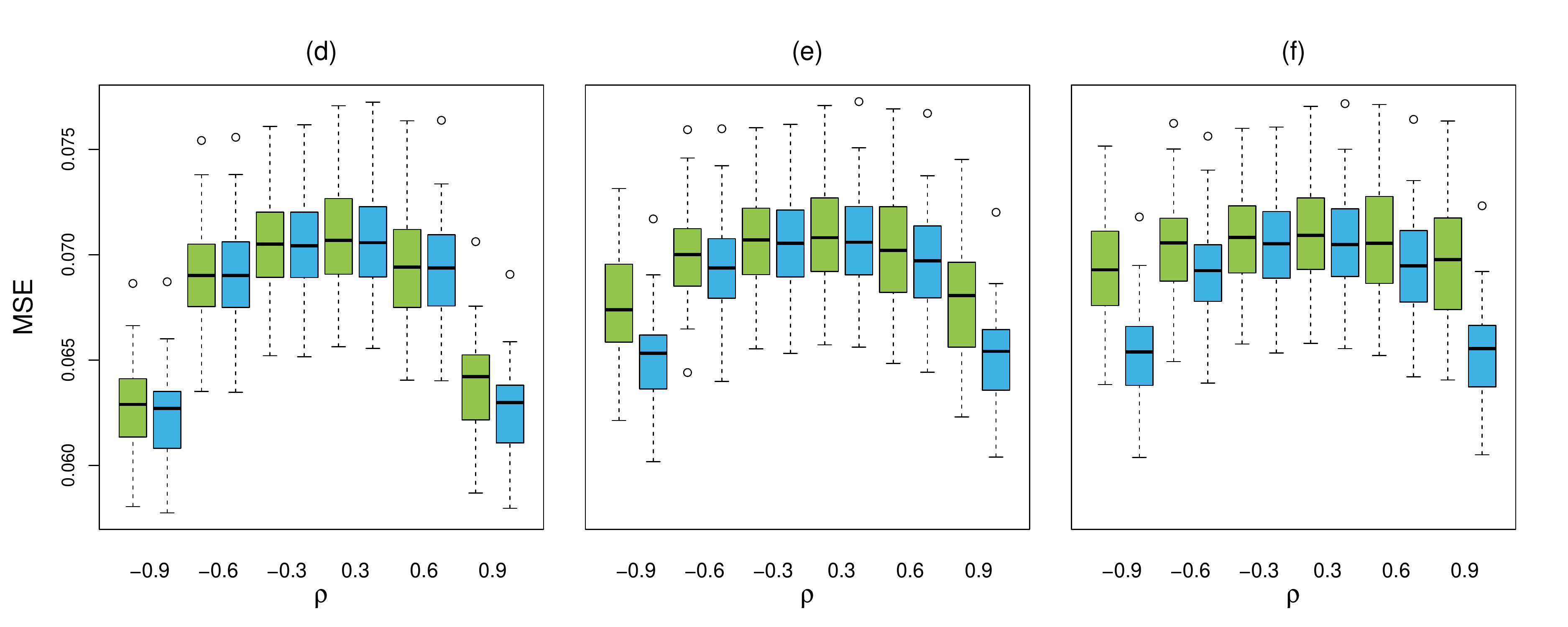}
	\caption{Boxplots of the MSEs under scenarios (d)-(f) when M2 (green) and M3 (blue) are fitted to data generated from M3 at different values of $\rho$.}
    \label{fig:simulation_study3}
\end{figure}

Additionally, we scrutinize the errors, summarized via boxplots in Figure~\ref{fig:simulation_study3}, when fitting M2 and M3 to data generated from M3. It can be seen that M2 does not predict as well as M3 in all cases but the difference in accuracy is the highest under scenario (f). This is because in scenario~(f), M2  severely overvalues the cross-covariances compared to the ones obtained by M3, resulting in a grave misspecification of spatio-temporal dependence between $Z_1$ and $Z_2$. All in all, in both experiments, the results demonstrate that neglecting the multiple advections phenomenon leads to substantial losses in prediction accuracy.

\section{Application} \label{sec:application}

Wind is recognized as a major driver of pollutant transport in the atmosphere. %The Modern-Era Retrospective Analysis for Research and Applications v.2 (MERRA-2) simulated black carbon concentrations over Saudi Arabia plotted in Figure~\ref{fig:data-fig1} demonstrate such transport behavior \citep{buchard2017merra}. 
Any physical model on pollutants factors in mechanisms of transport such as the wind fields at different levels of the troposphere \citep{kallos2007long},  jet streams, pyroconvective events, and boundary layer turbulence \citep{national2010global}. The transport behavior enables the propagation of suspended pollutants to locations far from the original source, which often results in higher pollutants concentrations on its path of transport. 

The modeling of pollutant concentrations is an active field of research in computational fluid dynamics (CFD). In that area of study, emphasis is placed on airflow motion and turbulence in determining pollutant concentrations at any location in space and time \citep{zhang2007comparison, katra2016modeling}. CFD involves tracking a large number of particles and employs highly specialized models, e.g., particle concentration equations coupled with momentum and turbulence equations, which need to be fed with various model input parameters \citep{knox1974numerical}. Lack of domain expert knowledge impedes adoption of these physically consistent models by non-experts in CFD.

The proliferation of pollutant measurements, which can be obtained from several sources such as ground stations, online databases (e.g., \href{https://goldsmr5.gesdisc.eosdis.nasa.gov/data/MERRA2/}{NASA Earthdata} website), satellite remote sensing, lidar networks, and other outdoor monitoring systems, has launched a new wave of statistical methodologies focused on modeling and predicting pollutant concentrations \citep{shao2011dust}. The list includes Bayesian models \citep{sahu2006spatio, calder2008dynamic}, generalized additive models \citep{paciorek2009practical, munir2013modeling}, geographically weighted regression \citep{van2016global}, stochastic partial differential equations \citep{cameletti2013spatio}, time series models \citep{goyal2006statistical}, and machine learning models \citep{mehdipour2018comparing}. These modeling approaches typically require only the historical measurements and other associated covariates. In the following section, we tackle the problem on hand by combining the strengths of Gaussian spatio-temporal geostatistical modeling and the concept of transport in CFD through the class of Lagrangian spatio-temporal models. 

\subsection{Saudi Arabia Black Carbon Data}\label{sec:application_data}

The data used in the present study were obtained from the \href{https://goldsmr5.gesdisc.eosdis.nasa.gov/data/MERRA2/}{NASA Global Modeling and Assimilation Office's} simulated MERRA-2 dataset containing pollutant concentration measurements. The values are given as mass mixing ratios, which measure the kilogram of pollutant per kilogram of air, with unit $kg / kg$. The list of accessible pollutant data includes black carbon, organic carbon, dust, and sea salt, from 1980 to present. We analyze and model only black carbon (BC) since it often contributes the highest to air pollution and has strong positive radiative forcing effect resulting to global warming \citep{schacht2019importance, takemura2019weak, rabito2020association}. %BC is a by-product of incomplete combustion of carbon-containing fuels and biomass burning \citep{zhang2020characteristics}. 

MERRA-2 BC concentrations are available at 3-hour intervals at 72 different pressure levels from the surface to 0.01 hPa on a regular grid with pixel size $0.5^{\circ} \times 0.625^{\circ}$. They were simulated using the Goddard Chemistry, Aerosol, Radiation, and Transport (GOCART) module (\textcolor{blue}{\url{https://acd-ext.gsfc.nasa.gov/People/Chin/gocartinfo.html}}), the Goddard Earth Observing System (GEOS-5) (\textcolor{blue}{\url{https://gmao.gsfc.nasa.gov/GEOS_systems}}), and atmospheric data assimilation system (ADAS) version 5.12.4 (\textcolor{blue}{\url{https://gmao.gsfc.nasa.gov/research/aerosol}}) \citep{randles2017merra}. GOCART is responsible for the sources, sinks, and chemistry aspect while GEOS-5 controls the earth science components of the simulations \citep{gelaro2017modern}.  

 \begin{figure}[t!]
 \centering
	\includegraphics[scale=0.27]{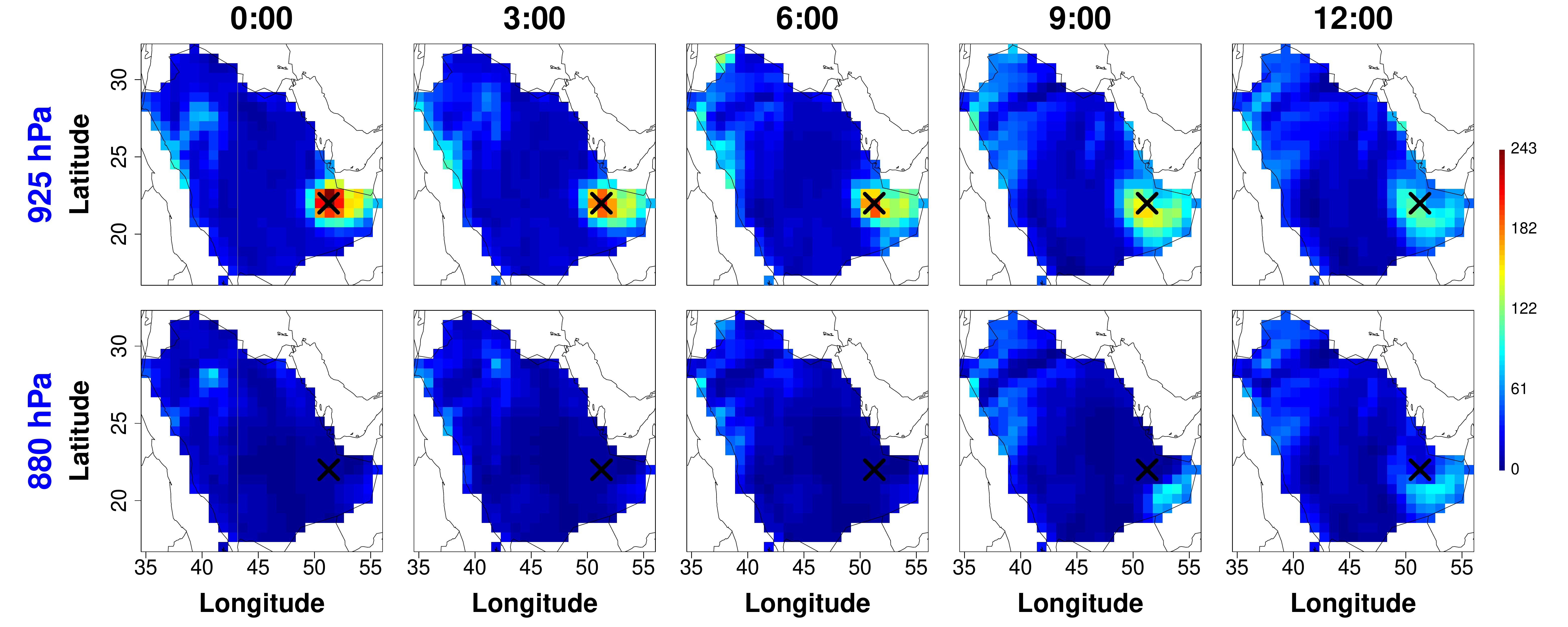}
	\caption{MERRA-2 black carbon concentrations in $10^{-12}$ $kg / kg$ at 3-hour increments from hours 0:00 to 12:00 on January 13, 2019 at two pressure levels. From the top row, the pressure levels are $\sim$ 0.7 and 1.2 kilometers away from the surface, respectively. A site is marked with ``$\times$" to aid detection of transport behavior.}
    \label{fig:data-fig1}
\end{figure}

To show the merits of the proposed multiple advections model, we build a bivariate dataset, $\mathbf{Y}(\mathbf{s}, t) = \{Y_1(\mathbf{s}, t), Y_2(\mathbf{s}, t)\}^{\top}$, using BC concentrations from pressure levels 925~hPa and 880~hPa, respectively. We hypothesize that the multiple advections model may outperform other alternative Gaussian spatio-temporal covariance function models since the components of the bivariate dataset come from different vertical regions of the atmosphere where advection behaviors may vary. Figure~\ref{fig:data-fig1} maps the BC concentrations over Saudi Arabia at five consecutive time periods on January 13, 2019. In this spatial domain, there are 550 sites such that the two nearest and the two farthest stations are$~$66 km and$~$2,204 km apart, respectively. Using the reference location as aid, a northward transport can be detected in both pressure levels. 

\subsection{The Lagrangian Spatio-Temporal Model}

In the simulated examples in the previous sections, such as those in Figure~\ref{fig:fig1}, the advection velocity vector, which is commonly associated with the wind vector in real environmental scenarios, is being identified as the main driver of the transport effect. In real data examples, however, several factors may be at play. The transport behavior observed in the BC data may be brought about by wet and dry depositions, time-varying sources, chemical reactions, and meteorological conditions, other than wind, over the area \citep{tanrikulu2009fine}. Hence, to model the bivariate spatio-temporal BC data, $\mathbf{Y}(\mathbf{s}, t)$, we assume that it has the following structure:
 \begin{equation} \label{eqn:real_data_linear_model}
Y_i(\mathbf{s}, t) = \mu_i (\mathbf{s},t) + Z_i(\mathbf{s} - \mathbf{V}_{ii} t),
\end{equation}
where $\mu_i(\mathbf{s}, t)$ is a mean function and $Z_i(\mathbf{s} - \mathbf{V}_{ii} t)$ is a zero-mean Gaussian spatio-temporal stationary process for variable $i$, $i = 1, 2$. The models we proposed in the paper are for the residual process $Z_i(\mathbf{s} - \mathbf{V}_{ii} t)$. Other time-varying factors that affect BC concentration levels are captured in the mean function $\mu_i(\mathbf{s}, t)$. Hence, we try to extract the stationary residual process $Z_i(\mathbf{s} - \mathbf{V}_{ii} t)$ such that there is no longer a recognizable trend in space and time after subtracting the mean function $\mu_i(\mathbf{s}, t)$ from $Y_i (\mathbf{s}, t)$. It may be that $\mu_i(\mathbf{s}, t)$ has some advection itself. To handle this advection behavior in the mean, we model $\mu_i(\mathbf{s}, t)$ as a linear combination of a set of covariates located at the same pressure level as $Y_i(\mathbf{s}, t)$ including organic carbon, dust, sulfur dioxide (SO$_2$), and sulphate aerosol (SO$_4$). These covariates are also suspended particles, thereby, containing information about the advection in the mean structure. Once $\hat{\mu}_i(\mathbf{s},t)$, an estimate of $\mu_i(\mathbf{s}, t)$, is obtained and is subtracted from $Y_i(\mathbf{s}, t)$, we are left with a residual bivariate spatio-temporal random field upon which we apply our proposed models. Any lingering transport behavior to this residual process we attribute to the random variable $\mathbf{V}_{ii}$ or the random wind vectors over the region.

\begin{figure}[t!]
 \centering
    \includegraphics[scale=0.27]{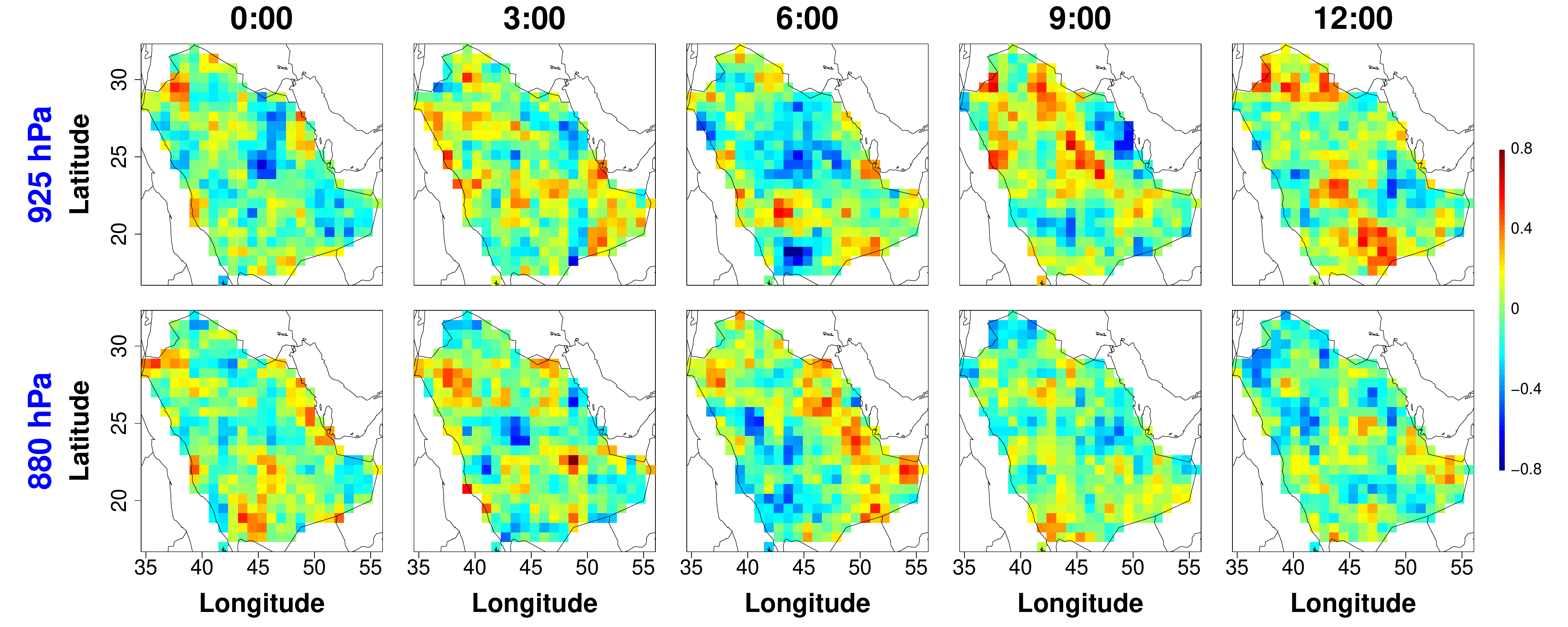}
	\caption{Residuals of the MERRA-2 black carbon concentrations in Figure~\ref{fig:data-fig1} in logarithmic scale. The bivariate random field appears to have a zero-mean advection, with no general direction of transport.}
	\label{fig:data-fig2}
\end{figure}

Furthermore, as an amalgamation of different complex processes, BC is a non-Gaussian and nonstationary process. In order to utilize the modeling framework discussed in the previous sections, Gaussianity and second-order stationarity of the BC data should be ensured. To obtain close to normally distributed measurements, we take the logarithm of the BC concentrations and pursue the spatio-temporal modeling on the logarithmic transformed dataset, i.e., $Y_i (\mathbf{s}, t) = \log \{ Y_i^0(\mathbf{s}, t) \}$, where $Y_i^0(\mathbf{s}, t)$ is the raw BC concentration at spatio-temporal location $(\mathbf{s}, t)$, for $i = 1, 2$; see \cite{paciorek2009practical}, \cite{sahu2012hierarchical}, and \cite{cameletti2013spatio} for similar treatments. To guarantee that we fit the models on spatio-temporal second-order stationary random fields, we identify consecutive time points such that, for each variable, the spatio-temporal residuals after removing the estimated mean via ordinary least squares, $\hat{\mu}_i(\mathbf{s}, t)$, passed the univariate test of stationarity in space and time devised in \cite{jun2012test}. In 2019, there are 139 non-overlapping spatio-temporal subsets that passed the test of stationarity. The maximum number of consecutive time points is six, which imputes a cross-covariance matrix of size 6,600 $\times$ 6,600. One of the spatio-temporal random fields in 2019 which passed the aforementioned test is the one shown in Figure~\ref{fig:data-fig1}. Its corresponding map of residuals after the estimated mean function under ordinary least squares was removed is displayed in Figure~\ref{fig:data-fig2}. %Despite removing $\hat{\mu}_i(\mathbf{s},t)$, where some advection information has been encoded through the covariates, we are still left with a residual bivariate random field with some recognizable transport behavior. 
The residual plots do not show a general direction of transport at both layers. Hence, it may be that the mean of $\mathbf{V}_{ii}$ is $(0, 0)^{\top}$ for $i = 1, 2$. 

 \begin{figure}[t!]
 \centering
	\includegraphics[width=\textwidth, height=19cm]{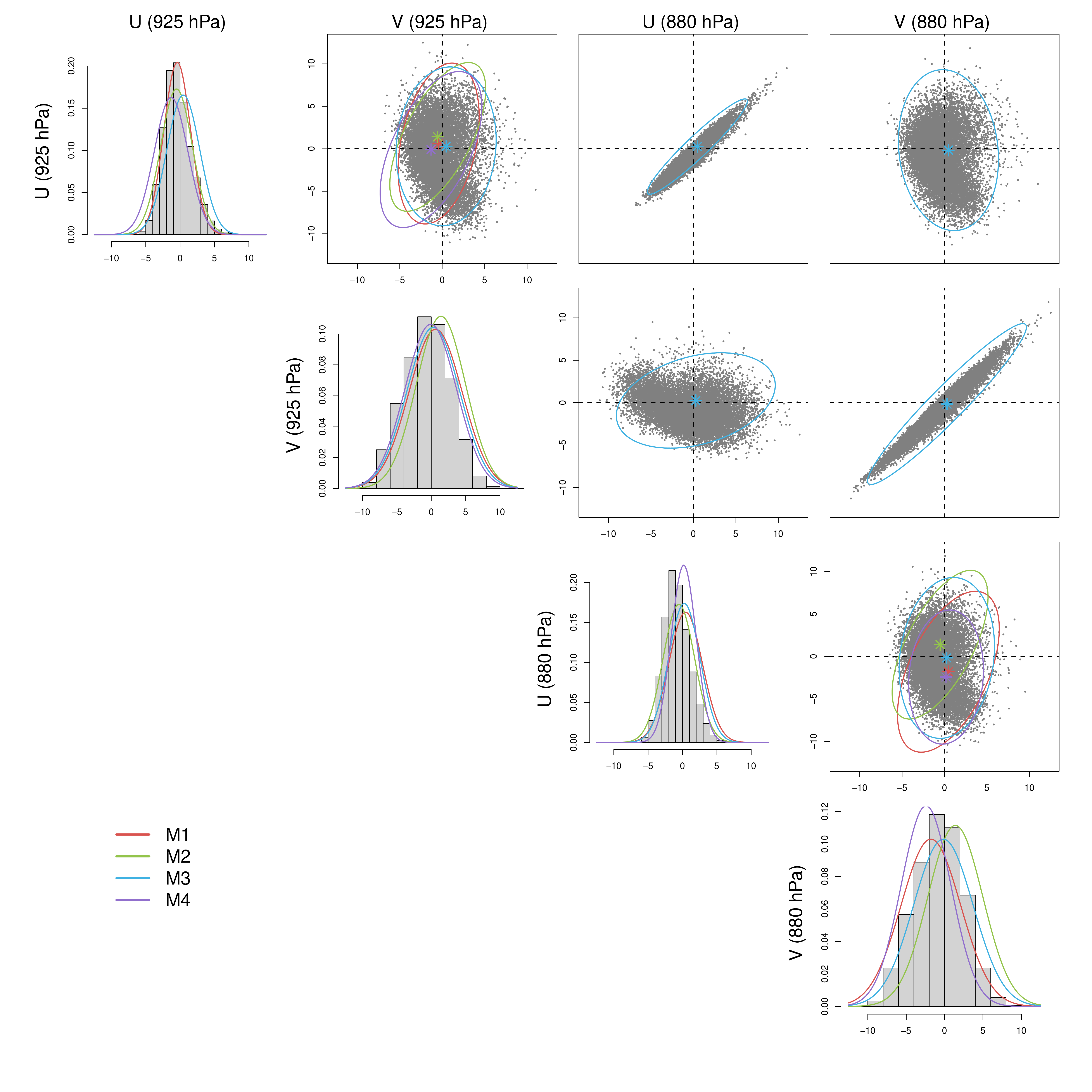}
	\caption{Empirical bivariate distributions of the MERRA-2 simulated wind vectors (in $m / s$) over Saudi Arabia. Here $U$ and $V$ denote the components of the wind vector along the $x$- and $y$-axis. Superimposed are the estimated distributions of the advection velocity vectors for the bivariate random field in Figure~\ref{fig:data-fig1} by different transport models M1 (red), M2 (green), M3 (blue), and M4 (purple), respectively. The lines in the bivariate plots enclose the 95\% probability region. Only M3 is applicable in the $2 \times 2$ panels from the top right corner.}
    \label{fig:data-fig3}
\end{figure}

In fact, the corresponding MERRA-2 simulated wind vectors over the spatio-temporal region under consideration support this zero-mean advection hypothesis. Figure~\ref{fig:data-fig3} plots the pairwise empirical bivariate distributions of the $U$ and $V$ components of the wind vectors in $m/s$ at the two pressure levels. These simulated wind vectors were used as inputs to simulate the BC concentrations \citep{randles2017merra, ukhov2020assessment}. From the plot, it can be seen that the wind vectors in 925~hPa and 880~hPa are centered at $(0, 0)^{\top}$. Moreover, the plot shows that the wind vectors at the two different layers are not identical otherwise we should see a straight line in the third plot in the first row and in the last plot in the second row. Instead, the $U$ and $V$ components at 925 hPa and 880 hPa appear to be highly correlated, with empirical correlation values of 0.9497 and 0.9787, respectively. Since the advections at the two layers are highly dependent, a single advection model may be thought to suffice. However, as mentioned in the simulation studies in Section~\ref{sec:simulation}, failing to acknowlege multiple advections behavior can reduce accuracy. Furthermore, the bivariate residuals display nonstationary colocated dependence. The empirical colocated correlations, $\hat{C}_{12} (\mathbf{0}; t, t) / \sqrt{\smash[b]{\hat{C}_{11}(\mathbf{0}, 0) \hat{C}_{22}(\mathbf{0}, 0) }}$, between the residuals at 925~hPa and 880 hPa are 0.56, 0.09, 0.06, 0.03, and $-0.06$ at $t = $ 0:00, 3:00, 6:00, 9:00, and 12:00, respectively. This resembles the property, which is well accommodated by the multiple advections model and not by the others, of colocated correlation approaching zero as time moves forward (see Figure~\ref{fig:fig2}). This indicates that different advection velocities may be affecting the two pressure levels and a multiple advections model may be the more suitable model for this dataset.

\subsection{Models}

To each of the spatio-temporal subsets that passed the aforementioned test, we fit five different spatio-temporal cross-covariance functions with bivariate parsimonious Mat\'{e}rn purely spatial margins, three of which were utilized in Section~\ref{sec:simulation}, namely, M1, M2, and M3, and the other two are as follows:
\begin{compactitem}
 
\item M4: bivariate Lagrangian spatio-temporal LMC in (\ref{eqn:spacetimelmc_expectation}) with uncorrelated latent univariate Lagrangian random fields $W_r(\mathbf{s}, t), r = 1, 2$; and

\item M5: bivariate non-Lagrangian fully symmetric spatio-temporal Gneiting-Mat\'{e}rn, i.e.,
\begin{equation*}\label{eqn:bivariategneitingmatern}
C_{ij}(\mathbf{h}, u) = \frac{\rho \sigma_{ii} \sigma_{jj} }{ (|u|^{2\xi} / \alpha + 1)^{b d / 2}}\mathcal{M}\left\{\frac{\|\mathbf{h}\|}{( |u|^{2\xi} / \alpha + 1)^{b/2}};a,\nu_{ij}\right\},\quad i,j = 1,2, 
\end{equation*}
where $\alpha>0$ and $\xi\in (0,1]$ describe the temporal range and smoothness, respectively \citep{bourotte2016flexible}. The parameter $b\in [0,1]$, also called the ``nonseparability parameter", represents the strength of the spatio-temporal interaction. 

\end{compactitem}

%Preliminary data analyses suggest that there is an anisotropic behavior in the marginal and cross-covariance structures. Hence, instead of evaluating the models at $\mathbf{h}$, we use $\mathbf{R} \mathbf{h}$, where $\mathbf{R} = \begin{pmatrix}
%R_1 \cos R_3 & R_1 \sin R_3 \\
%-R_2 \sin R_3 & R_2 \cos R_3 \\
%\end{pmatrix}$ is the anisotropy matrix parameterized by the anisotropy scale parameters, $R_1$ and $R_2$, and the anisotropy angle parameter, $R_3$; see \cite{paciorek2006spatial} and \cite{hewer2017matern} for discussions on this modeling approach and more general parameterizations of $\mathbf{R}$. 

\subsection{Results and Discussions}

The parameters of the Lagrangian models M1, M2, M3, and M4 are estimated via multi-step REML, as presented in Section~\ref{sec:estimation}, while those of the non-Lagrangian model M5 are estimated jointly. In each subset, we use all spatio-temporal measurements except those at the last time point to fit the models. Once the parameters were obtained, we predict the measurements at all spatial locations at the last time point. Estimation and prediction routines are implemented in parallel using the high performance libraries, namely, BLACS, PBLAS, and ScaLAPACK, through the \texttt{R} package \texttt{pbdBASE}, which provides a set of wrappers for those libraries \citep{schmidt2020guide}. The codes are run on a cluster using 3 nodes of 2.6 GHz Intel Skylake with 40 cores per node and 350 GB memory.

 \begin{figure}[t]
 \centering
	\includegraphics[scale=0.5]{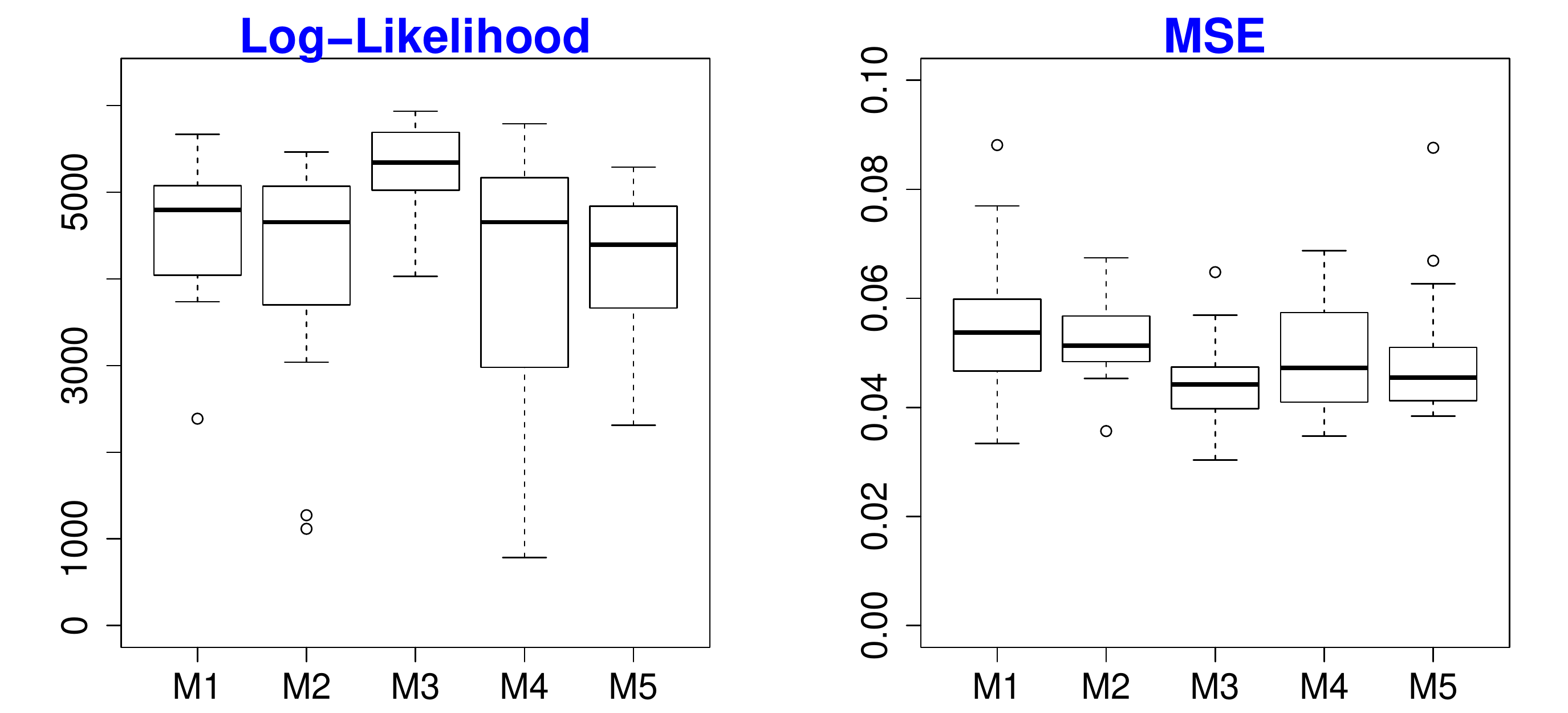}
	\caption{The log-likelihood and MSE values recovered from fitting the five models on all 139 non-overlapping bivariate spatio-temporal second-order stationary log BC residuals.}
    \label{fig:loglik-mse}
\end{figure}

Figure~\ref{fig:loglik-mse} shows the values of the in-sample and out-of-sample scores, namely, log-likelihood and MSE, recovered from fitting the five models on all 139 non-overlapping bivariate spatio-temporal second-order stationary log BC residuals. It can be seen from the figure that the multiple advections model M3 emerged as the best performing model in both metrics. %This is not surprising as the empirical colocated correlation values, $\hat{C}_{12} (\mathbf{0}; t, t) / \sqrt{\smash[b]{\hat{C}_{11}(\mathbf{0}, 0) \hat{C}_{22}(\mathbf{0}, 0)}}$, obtained from the BC data are nonstationary. In January 13, 2019, for instance, the empirical colocated correlations between the residuals at 925~hPa and 880~hPa are 0.64, 0.12, -0.28, 0.26, and 0.14 at $t = $ 0:00, 3:00, 6:00, 9:00, and 12:00, respectively. 
Moreover, the bivariate models, M2-M5, outperform the univariate model, M1, in terms of prediction. This hints that BC at different layers of the atmosphere have substantial dependence which should be taken into account to obtain more accurate predictions. Lastly, although M3 clearly is the best model in terms of the log-likelihood, the prediction performance of M5 does not lag that far behind. The danger, however, in settling with M5 is that fitting a non-multiple advections model on random fields with possible multiple advections can lead to erroneous parameter estimates as illustrated by Figure~\ref{fig:simulation_study2} in Section~\ref{sec:simulation}. This could ultimately distort the insights regarding the true nature of dependence between the variables.

\begin{table}[t!]
\centering
\caption{Median in-sample (log-likelihood, AIC$^*$, and BIC$^*$) scores, out-of-sample (MSE) scores, and computation times for the 139 non-overlapping bivariate spatio-temporal second-order stationary log BC residuals. The best scores are given in bold.}
\scalebox{0.9}{
\begin{tabular}{cccccccccccccc}
    \toprule
    \multirow{2}{*}{}
      &   \multicolumn{3}{c}{In-Sample} &  Out-of-Sample &  No. of & Average Computation \\
     Model & log-likelihood & AIC$^*$ & BIC$^*$ & MSE & Parameters & Time (hours) \\
    \cmidrule(lr){2-4}  \cmidrule(lr){5-5} \cmidrule(lr){6-6} \cmidrule(lr){7-7}
    M1 & 4,845 & $-$9,657 & $-$9,566 & 0.0556 & 16 & 0.26 \\
    M2 & 4,652 & $-$9,281 & $-$9,211 & 0.0513 & 11 & 0.98  \\
    M3 & \bf{5,343} & \bf{$-$10,647} & \bf{$-$10,519} & \bf{0.0442} & 20 & 1.35 \\
    M4 & 4,985 & $-$9,934 & $-$9,819 & 0.0460 & 18 & 1.17 \\
    M5 & 4,458 & $-$8,895 & $-$8,831 & 0.0454 & 9 & 0.57 \\
    \bottomrule
  \end{tabular}
  }
%\caption*{} 
\label{tab:results}
\end{table}

Table~\ref{tab:results} reports the median values of the boxplots in Figure~\ref{fig:loglik-mse} including the median Akaike (AIC$^*$) and Bayesian information criteria (BIC$^*$). Here AIC$^*$ and BIC$^*$ are the multi-step versions of the classical AIC and BIC, i.e., AIC$^* = - 2 l_{\text{REML}}( \boldsymbol{\hat{\theta}}_{-\boldsymbol{\Sigma}_{\boldsymbol{\mathcal{V}}}},  \boldsymbol{\hat{\theta}}_{\boldsymbol{\Sigma}_{\boldsymbol{\mathcal{V}}}}, \hat{\boldsymbol{\beta}}_{\text{GLS}}) + 2 q$ and BIC$^* = - 2 l_{\text{REML}}( \boldsymbol{\hat{\theta}}_{-\boldsymbol{\Sigma}_{\boldsymbol{\mathcal{V}}}}, \boldsymbol{\hat{\theta}}_{\boldsymbol{\Sigma}_{\boldsymbol{\mathcal{V}}}}, \hat{\boldsymbol{\beta}}_{\text{GLS}}) + q \log n$, where $l_{\text{REML}}( \boldsymbol{\hat{\theta}}_{-\boldsymbol{\Sigma}_{\boldsymbol{\mathcal{V}}}},  \boldsymbol{\hat{\theta}}_{\boldsymbol{\Sigma}_{\boldsymbol{\mathcal{V}}}}, \hat{\boldsymbol{\beta}}_{\text{GLS}})$ is the value of the log-likelihood function at the second estimation step with parameter estimates $ \boldsymbol{\hat{\theta}}_{\boldsymbol{\Sigma}_{\boldsymbol{\mathcal{V}}}}$ while fixing the parameters $ \boldsymbol{\hat{\theta}}_{-\boldsymbol{\Sigma}_{\boldsymbol{\mathcal{V}}}}$ obtained at the first estimation step. For the one step estimation procedure of M5, AIC$^* = - 2 l_{\text{REML}}( \boldsymbol{\hat{\Theta}}, \hat{\boldsymbol{\beta}}_{\text{GLS}}) + 2 q$ and BIC$^* = - 2 l_{\text{REML}}( \boldsymbol{\hat{\Theta}}, \hat{\boldsymbol{\beta}}_{\text{GLS}}) + q \log n$. The number of parameters and average computation time of each model are also indicated in Table~\ref{tab:results}. From the table, it can be seen that the AIC$^*$ and BIC$^*$ support the usage of M3 even though the model has the most number of parameters. Additionally, the estimates associated with the transport phenomenon are validated against the wind vectors observed in the spatio-temporal domain under study. In Figure~\ref{fig:data-fig3}, the estimated distribution of the advection velocity vectors in the transport models M1-M4 are superimposed on the MERRA-2 simulated wind vectors over Saudi Arabia. From the figure, it can be seen that the advection parameters in all four transport models adequately capture the empirical distribution, with actual empirical coverages of 92\% (M1), 86\% (M2), 87\% (M3), and 84\% (M4), for the wind vectors at 925~hPa, and 83\% (M1), 85\% (M2), 86\% (M3), and 94\% (M4), for the wind vectors at 880~hPa.

\section{Conclusion} \label{sec:discussion}

We successfully pursued the development of Lagrangian spatio-temporal cross-covariance
functions with multiple advections. The proposed framework offers a suite of 
data-driven models which is more flexible and realistic. We also outlined an estimation procedure to get sensible estimates of all parameters. We showed through numerical experiments involving bivariate spatio-temporal random fields how failing to account for multiple advections produces poor predictions and how utilizing simpler univariate Lagrangian spatio-temporal models cannot capitalize on the purely spatial variable dependence. With wind recognized as the main driver of the transport behavior, the proposed models, along with another benchmark non-Lagrangian spatio-temporal model, were tested on pollutant data over Saudi Arabia. Indeed, a model under our proposed class emerged as the best model and its estimate of the multivariate distribution of the advection velocities checks out with the prevailing behavior of wind over the area.

\section*{Appendix}

\noindent {\bf Proof of Theorem \ref{theorem1}}
Let $\boldsymbol{\lambda}_l\in\mathbb{R}^{p}, \;l=1,\ldots,n.$ Then:
\begin{eqnarray*}
\sum_{l=1}^{n}\sum_{r=1}^{n}\boldsymbol{\lambda}_l^{\top} \mathbf{C}(\mathbf{s}_l, \mathbf{s}_r; t_l, t_r) \boldsymbol{\lambda}&=&\sum_{l=1}^{n}\sum_{r=1}^{n}\boldsymbol{\lambda}_l^{\top} \text{E}_{\boldsymbol{\mathcal{V}}} \left( \left[C_{ij}^S\left\{(\mathbf{s}_l - \mathbf{V}_{ii} t_l)-(\mathbf{s}_r - \mathbf{V}_{jj} t_r )\right\}\right]_{i, j = 1}^{p} \right) \boldsymbol{\lambda}_r \\
&=& \text{E}_{\boldsymbol{\mathcal{V}}} \left[ \sum_{l=1}^{n}\sum_{r=1}^{n}\boldsymbol{\lambda}_l^{\top} \{ C_{ij}^S (\mathbf{s}_l - \mathbf{s}_r - \mathbf{V}_{ii} t_l + \mathbf{V}_{jj} t_r ) \}_{i, j = 1}^{p} \boldsymbol{\lambda}_r \right] \geq 0 
\end{eqnarray*}
for all $n\in\mathbb{Z}^+$ and $\left\{(\mathbf{s}_1,t_1),\ldots,(\mathbf{s}_n,t_n)\right\} \in \mathbb{R}^d\times \mathbb{R}$. The last inequality follows from the assumption that $\mathbf{C}^S$ is a valid purely spatial matrix-valued stationary cross-covariance function with variable asymmetry on $\mathbb{R}^d$; see \cite{li2011approach}, \cite{genton2015cross}, and \cite{qadir2020flexible} for discussions on this class of cross-covariance functions. \hfill $\Box$ \\

\vspace{-.7cm}

\noindent {\bf Proof of Theorem \ref{theorem_multiple_advec_nonfrozen}} See Supplementary Materials.

\noindent {\bf Proof of Theorem \ref{variable_specific_advec}} The validity is guaranteed by the dimension expansion approach in \cite{bornn2012modeling}.  \hfill $\Box$

\noindent {\bf Proof of Theorem \ref{lmc_multiple}} The validity of (\ref{eqn:spacetimelmc_expectation}) is established as it is the resulting spatio-temporal cross-covariance function of the process in (\ref{eqn:lmcprocess}).  \hfill $\Box$ 

\bigskip

\noindent {\bf Derivation of Equation(\ref{eqn:locally_stationary})}
\begin{small}
\begin{eqnarray*}
C_{ij}\left(\mathbf{h}-\mathbf{V}_{ii}t_l+\mathbf{V}_{jj}t_r\right) & = & C_{ij}\left\{\mathbf{h}-\mathbf{V}_{ii}\left(m+\frac{u}{2}\right)+\mathbf{V}_{jj}\left(m-\frac{u}{2}\right)\right\} = C_{ij}\left\{\mathbf{h}-\overline{\mathbf{V}}_{ij}u+\left(\mathbf{V}_{jj}-\mathbf{V}_{ii}\right)m\right\},
\end{eqnarray*}
\end{small}
where $u=t_l-t_r$, $m=\frac{t_l+t_r}{2}$, and $\overline{\mathbf{V}}_{ij}=\frac{\mathbf{V}_{ii} + \mathbf{V}_{jj}}{2}$. \hfill $\Box$

\bigskip

\noindent {\bf Codes}

\noindent The codes can be found in this website: \textcolor{blue}{\url{https://github.com/geostatistech/multiple-advections}}

\baselineskip=20pt
\bibliographystyle{apalike}
\bibliography{main}
\end{document}